\documentclass[prc,aps,preprint,tightenlines,nofootinbib]{revtex4-1}

\usepackage{appendix}
\usepackage{graphicx}
\usepackage{slashed}
\usepackage{amsmath}
\usepackage{hyperref}
\hypersetup{colorlinks=true,citecolor=blue,linkcolor=blue,urlcolor=blue}

\begin{document}
\title {Probing Vector Mesons in Deuteron Break-up Reactions}
\author{Adam J. Freese and Misak M. Sargsian}
\affiliation{Department of Physics, Florida International University, Miami, FL 33199}
\date{\today}

\begin{abstract}
We study vector meson photoproduction from the deuteron at high momentum transfer,
accompanied by break-up of the deuteron into a proton and neutron. The large $-t$ involved
allows one of the nucleons to be identified as struck, and the other as a spectator to the
$\gamma N\rightarrow VN$ subprocess. Corrections to the plane wave impulse approximation
involve final state interactions (FSIs) between the struck nucleon or the vector meson,
either of which is energetic, with the slow spectator nucleon. In this regime, the eikonal
approximation is valid, so is employed to calculate the cross-section for the reaction.
Due to the high-energy nature of the FSIs, the maxima of the rescatterings are located at
nearly transverse directions of the fast hadrons. This results in two peaks in the angular
distribution of the spectator nucleon, each corresponding to either the $V$-$N$ or the
$p$-$n$ rescattering. The $V$-$N$ peak provides a new means of probing the $V$-$N$
interaction. This is checked for near-threshold $\phi$ and $J/\Psi$ photoproduction
reactions which demonstrate that the $V$-$N$ peak can be used to extract the largely
unknown amplitudes of $\phi$-$N$ and $J/\Psi$-$N$ interactions.

Two additional phenomena are observed when extending the calculation of $J/\Psi$
photoproduction to the sub-threshold and high-energy domains. In the first case we observe
overall suppression of FSI effects due to a restricted phase space for sub-threshold
production in the rescattering amplitude. In the second, we observe cancellation of the
$V$-$N$ rescattering amplitudes for vector mesons produced off of different nucleons in
the deuteron.
\end{abstract}
\maketitle
 
\section{Introduction}
\label{sec:intro}

Photoduction of vector mesons from nuclei with subsequent rescattering is the main
approach in extracting the cross-sections of $V$-$N$ interactions that does not require
the assumption of vector meson dominance\cite{Bauer:1977iq}. To enhance the measured
cross-sections, experiments have been carried out on medium to heavy nuclei, where the
extraction of the $V$-$N$ scattering amplitude is based on the Glauber theory of multiple
hadronic interactions\cite{Glauber:1970jm}. In several instances a deuteron target has
been used, as it provides higher theoretical accuracy in the extraction of the
$VN\rightarrow VN$ amplitude\cite{Bauer:1977iq}. In the case of the deuteron target, one
method is to study the coherent photoproduction at large
$-t\ge 0.8~\mathrm{GeV}^2$\cite{Anderson:1971ar,Frankfurt:1997ss,Rogers:2005bt,Mibe:2007aa},
for which the $\gamma d\rightarrow Vd^\prime$ cross-section is primarily due to
rescattering of the vector meson (first produced on a struck nucleon) off the spectator
nucleon. This property has been used for probing both
$\rho$-$N$\cite{Anderson:1971ar,Frankfurt:1997ss} and
$\phi$-$N$\cite{Mibe:2007aa,Rogers:2005bt} cross-sections.
In these studies, calculations involving the elastic $V$-$N$ scattering amplitude are fit
to experimental data to extract the total $V$-$N$ cross-section, the slope factor $B_{VN}$, and the
ratio of the real and imaginary parts of the amplitude, $\alpha_{VN}$. However, this
procedure has had limited success, since these parameters cannot be unambiguously
disentangled by just the measurement of cross-sections. This situation can be improved by
considering polarization observables of the coherent reaction\cite{Frankfurt:1997ss}, but
the experimental realization of polarization processes is more difficult.

In this work we suggest a new approach for measuring $V$-$N$ scattering cross section by
considering vector meson photoproduction from the deuteron with subsequent break-up of
the deuteron into a proton and neutron. We focus on kinematics with a large momentum
transfer, for which the fast, ``struck'' nucleon can be unambiguously identified apart
from the slow, ``recoil'' nucleon which is a spectator to the $\gamma N\rightarrow VN$
subprocess in the plane wave impulse approximation (PWIA). In this case, the final
state interactions are characterized by a rescattering of either of the fast outgoing
hadrons---the vector meson or the struck nucleon---with the spectator nucleon. By
selecting the momentum and the direction of the spectator nucleon, one can increase the
sensitivity of the reaction to small angle $V$-$N$ scattering, thereby probing the
parameters of the $VN\rightarrow VN$ amplitude.

In addition to allowing identification of the struck and spectator nucleons, the large
$-t$ establishes the applicability of the eikonal approximation to the hadronic
rescatterings. In particular, we use the generalized eikonal approximation
(GEA)\cite{Frankfurt:1994kt,Frankfurt:1996xx,Sargsian:2001ax,CiofidegliAtti:2004jg}, which
was developed to account for the finite target and recoil nucleon momenta neglected in the
conventional Glauber approximation. GEA has been successfully applied to high energy
electrodisintegration of the deuteron\cite{Sargsian:2009hf}, semi-inclusive deep inelastic
scattering\cite{Cosyn:2010ux}, and coherent photoproduction of
$\rho$\cite{Frankfurt:1997ss} and $\phi$\cite{Rogers:2005bt} mesons from the deuteron.

The most relevant of these to this paper is deuteron electrodisintegration, i.e. 
$ed\rightarrow e^\prime pn$, in which one of the nucleons is struck and the other is
a spectator in PWIA. In this case the eikonal approximation predicts a clear $p$-$n$
rescattering peak when the differential cross-section is plotted against the spectator
angle for spectator momenta larger than $400~\mathrm{GeV/c}$. This feature was
experimentally confirmed in high momentum transfer deuteron break-up
reactions\cite{Egiyan:2007qj,Boeglin:2011mt}.

In the reaction considered in this work,
we see a new phenomenon, in which the presence of two energetic hadrons 
to rescatter from the slow spectator nucleon gives the possibility of observing two peaks,
one associated with the $V$-$N$ and the other with the $p$-$n$ rescattering. One of our
goals is to investigate whether the $V$-$N$ peak can be used to study the properties of
the $V$-$N$ interaction amplitudes.

The article is organized as follows: In section \ref{sec:reaction} we consider the
kinematics of the reaction and justify the application of the generalized eikonal
approximation. In section \ref{sec:calc} we calculate the amplitude for the reaction
within the virtual nucleon approximation, in which the rescatterings are calculated based
on GEA. In section \ref{sec:num} we present numerical estimates for the break-up
reaction, with photoproduction of $\phi(1020)$ and of $J/\Psi$.
The appendix provides details of the calculation of the double rescattering amplitude.

\section{Reaction, kinematics and validity of eikonal approximation}
\label{sec:reaction}

We study large momentum transfer ($-t\ge 1~\mathrm{GeV}^2$) photoproduction
from the deuteron accompanied by deuteron break-up:
\begin{equation}
	\gamma + d \rightarrow V + p + n.
\label{eqn:reaction}
\end{equation}
Our main goal is to investigate the feasibility of using this reaction to study the
amplitude of $V$-$N$ scattering. This goal can be achieved only if it is possible to
unambiguously isolate and reliably calculate the contribution of $V$-$N$ rescattering to
the total cross-section to this reaction. 

\begin{figure}[hpt]
	\centering\includegraphics[scale=0.5]{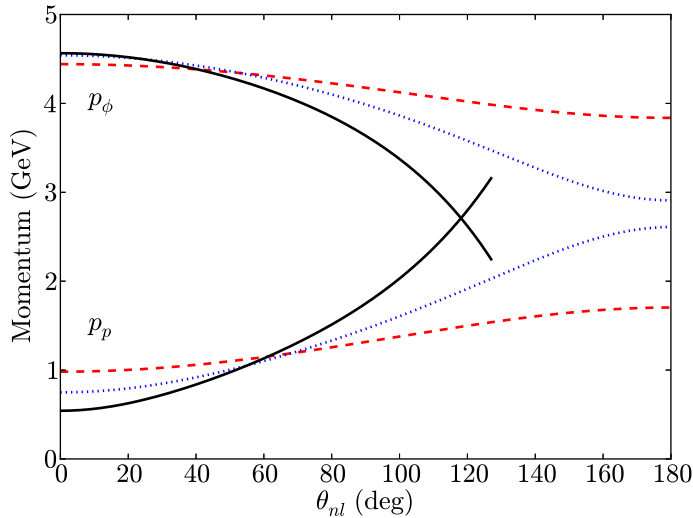}
	\caption{
		(Color online)
		Dependence of the momenta of struck nucleon and $\phi$ meson on the momentum and
		direction of the recoil nucleon, at $q_0=5~\mathrm{GeV}$ and
		$t=-1.2~\mathrm{GeV}^2$, for the reaction (\ref{eqn:reaction}).
		The dashed (red), dotted (blue), and solid (black) curves correspond respectively
		to recoil momenta of $200$, $400$, and $600~\mathrm{MeV}$.
		The angle is defined with respect to the momentum transfer $\mathbf{l}$ defined in
		Eq.~(\ref{eqn:l}).
	}
	\label{fig:kinphi}
\end{figure}

The experience of recent years\cite{Sargsian:2009hf,Frankfurt:1997ss,Jeschonnek:2008zg,
	CiofidegliAtti:2004jg,CiofidegliAtti:2000xj,Laget:2004sm,Mibe:2007aa,Egiyan:2007qj,
	Boeglin:2011mt}
demonstrates that such an isolation can be achieved in the eikonal regime, where the
momenta of the rescattered fast hadrons are on the order of a GeV/c and above. The main
advantage of the eikonal approximation is that the part of the $\gamma d\rightarrow Vpn$
amplitude corresponding to the pole values of the intermediate hadronic propagators is
expressed through the on-shell hadronic rescattering amplitudes, which can be related to
the experimental values of the total $V$-$N$ and $p$-$n$ cross-sections. The contribution
from the principal value parts is expressed through half-off-shell hadronic amplitudes. In
situations where the latter contributions are small, the eikonal regime provides the
possibility of extracting on-mass-shell amplitudes of $V$-$N$ and $p$-$n$ elastic
scattering in kinematics dominated by final state interactions.

To establish the eikonal regime for the reaction (\ref{eqn:reaction}), we concentrate
on kinematics where the subprocess $\gamma N\rightarrow VN$ has a large momentum
transfer, and where just one of the nucleons emerges with a momentum
$\gtrsim 1~\mathrm{GeV/c}$. For definiteness we consider this to be the proton, so that
the final state of the reaction consists of two energetic hadrons (the vector meson and
the proton) and a slow recoil neutron. In such a situation, the final state interactions
can be calculated within the eikonal approximation.

\begin{figure}[hpt]
	\centering\includegraphics[scale=0.5]{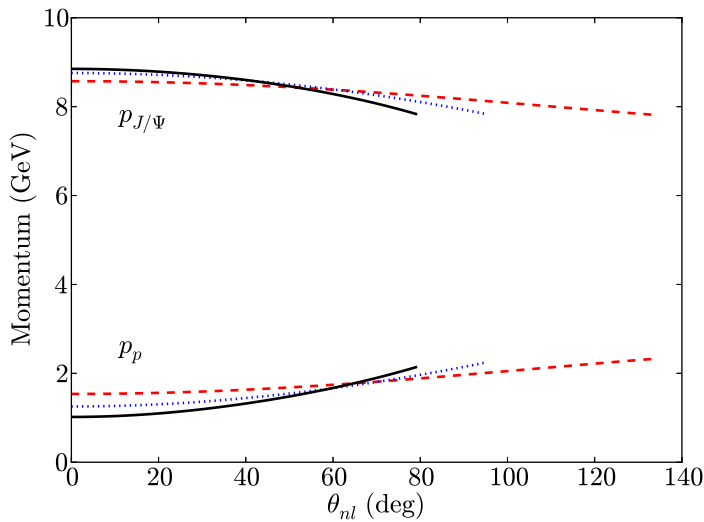}
	\caption{
		(Color online)
		Dependence of the momenta of struck nucleon and $J/\Psi$ meson on the momentum and
		direction of the recoil nucleon, given $q_0=10~\mathrm{GeV}$ and
		$t=-2.23~\mathrm{GeV}^2$, for the reaction (\ref{eqn:reaction}).
		The curves have the same meaning as in Fig.~\ref{fig:kinphi}.
	}
	\label{fig:kinjpsi}
\end{figure}

For light vector mesons, such a regime can be established if one considers kinematics
where the momentum transfer at the $\gamma N\rightarrow VN$ vertex is sufficiently large,
namely $-t\ge 1~\mathrm{GeV}^2$. As Fig.~\ref{fig:kinphi} demonstrates, for
$t=-1.2~\mathrm{GeV}^2$ both the $\phi$ meson and the proton are sufficiently energetic
for different values of the neutron momentum. The photon energy in this figure is
$5~\mathrm{GeV}$ and the value of $t$ is chosen on the proven feasibility of probing
$\phi$ photoproduction at this $t$ value in coherent processes at Jefferson
Lab\cite{Mibe:2007aa}. Hereafter, we represent the neutron direction using the angle it
makes with respect to the direction of the momentum transfer.

For heavier mesons such as $J/\Psi$, the eikonal regime is established naturally near
threshold kinematics due to the large value of $-t_{\mathrm{thr.}}$ (as calculated for a
stationary proton target). This can be seen in Fig.~\ref{fig:kinjpsi}, where the momenta
of the proton and $J/\Psi$ meson are presented for different values of the neutron
momentum, given a near-threshold photon energy of $10~\mathrm{GeV}$ (the threshold being
$8.2~\mathrm{GeV}$), and $t=t_{\mathrm{thr.}}$. 

Provided the eikonal regime is established, we can use a diagrammatic approach to
calculate the reaction amplitude, which is expanded in the number of rescatterings
(see e.g.~\cite{Sargsian:2001ax}). In such an expansion, we will have only three
categories of scattering amplitudes: a 0th order term corresponding to the plane wave
impulse approximation, and 1st and 2nd order terms corresponding to terms with one or two
hadronic rescatterings.

\section{Calculation of the scattering amplitude and cross section}
\label{sec:calc}

We denote the four-momenta of the deuteron, photon, vector meson, proton, and neutron 
as $p_d=(E_d,\mathbf p_d)$, $q=(q_0,\mathbf q)$, $p_V=(E_V,\mathbf p_{V})$
$p_{p}=(E_{p},\mathbf p_{p})$ and $p_{n} = (E_{n},\mathbf p_{n})$, respectively. We also
denote the ``initial'' momenta of the proton and neutron within the deuteron (prior to
vector meson photoproduction) as $p^\prime_{p}$ and $p^\prime_{n}$, respectively. The
four-momentum transfer to the deuteron is defined as
\begin{equation}
	l^\mu = (l_0,\mathbf l) \equiv (q_0-E_V,\mathbf q-\mathbf p_V)
	\label{eqn:l}
\end{equation}
with $t=l^2$. As was mentioned above, the direction of the outgoing neutron is defined by
the angle it makes with the momentum transfer $\mathbf l$, and is denoted $\theta_{nl}$.

The general expression for the differential cross-section of the reaction
(\ref{eqn:reaction}) is:
\begin{equation}
	d\sigma
	=
	\frac{1}{4(q\cdot p_d)}
	\overline{\mid\mathcal{M}\mid^2}
	(2\pi)^4 \delta^4(q + p_d - p_V - p_{p} - p_{n}) 
	\frac{d^3\mathbf p_V}{(2\pi)^3 2E_V}
	\frac{d^3\mathbf p_{p}}{(2\pi)^3 2E_{p}}
	\frac{d^3\mathbf p_{n}}{(2\pi)^3 2E_{n}},
	\label{eqn:crx}
\end{equation}
where we sum over the final and average over the initial particle polarizations. In our
approach, the Feynman amplitude $\mathcal{M}$  is expanded in terms of the 
number of hadronic rescatterings (see e.g.~\cite{Sargsian:2001ax})
\begin{equation}
	\mathcal{M} = \mathcal{M}_0 + \mathcal{M}_1 + \mathcal{M}_2,
	\label{M012}
\end{equation}
where, $\mathcal{M}_0$, $\mathcal{M}_1$ and $\mathcal{M}_2$ correspond respectively to
PWIA and the single and double rescattering amplitudes. These amplitudes will be
calculated within the virtual nucleon approximation (VNA), in which only the $pn$
component of the deuteron wave-function and only the positive-energy pole of each
nucleon's propagator is considered. This allows us to relate the covariant
$d\rightarrow pn$ transition amplitude to the deuteron wave-function with a virtual struck
and an on-shell spectator nucleon (see e.g.~\cite{Buck:1979ff}). Because the non-nucleonic
components of the deuteron and the negative-energy poles of the nucleon propagators are
neglected, the deuteron wave-fuction does not saturate the momentum sum rule. However, the
relativistic normalization of the wave-function is fixed by baryon number conservation
(see e.g. \cite{Frankfurt:1976gz}). Previous estimates\cite{Sargsian:2009hf} demonstrate
that this approximation works reasonably well for deuteron internal momenta 
$\lesssim700~\mathrm{MeV/c}$, which will be considered as a kinematical limit for the
present calculations.
 
\subsection{Plane Wave Impulse Approximation}
\label{sec:PWIA}

\begin{figure}[hpt]
	\centering\includegraphics[scale=0.5]{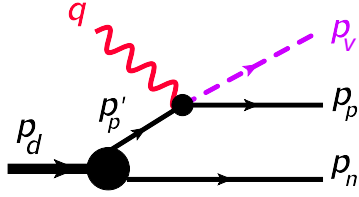}
	\caption{
		(Color online)
		Feynman diagram for PWIA scattering.
	}
	\label{fig:pwia}
\end{figure}

In plane wave impulse approximation (PWIA), we neglect the final state interactions
between outgoing hadrons, treating them as plane waves (Fig.~\ref{fig:pwia}). The
covariant scattering amplitude $\mathcal{M}_0$ within PWIA can be calculated based on
effective Feynman diagram rules (see e.g. \cite{Sargsian:2001ax}) which give:
\begin{align}
	\mathcal{M}^{(\lambda_{V}\lambda_{p}\lambda_{n};\lambda_\gamma\lambda_d)}_0
		&=
		-\bar{u}^{(\lambda_{p})}(\mathbf p_{p})
		\bar{u}^{(\lambda_{n})}(\mathbf p_{n})
		{\phi^\dagger}^{(\lambda_V)}_\nu(\mathbf p_V)
		\Gamma^{\mu\nu}_{\gamma N\rightarrow VN}
		\epsilon_\mu^{(\lambda_\gamma)}(q)
	\notag \\ &\times
		\frac{\slashed{p}^\prime_p+m_N}{p^{\prime 2}_p-m_N^2+i\epsilon}
		\Gamma_{dpn}\chi_d^{(\lambda_d)}(\mathbf p_d)
	.
	\label{eqn:M0_a}
\end{align}
Here, $\Gamma_{dpn}$ and $\Gamma^{\mu\nu}_{\gamma N\rightarrow VN}$ are the covariant
vertices for the transitions $d\rightarrow pn$ and $\gamma N\rightarrow VN$. The spin
wave-functions of the deuteron, nucleons, photon and vector meson are denoted
$\chi_d$, $u$, $\epsilon_\mu$, and $\phi_\nu$, respectively. The spin degree of freedom 
of each particle is identified by a superscript in parentheses.

As was discussed above, we evaluate Eq.~(\ref{eqn:M0_a}) by considering only the
positive-energy pole in the bound nucleon propagator. Then, in accordance with VNA we use
the approximate completeness relation
$
	\slashed{p}^\prime_p+m_N
	\approx
	\sum_{\lambda^\prime_{p}}
	u^{(\lambda^\prime_{p})}(\mathbf p^\prime_{p})
	\bar{u}^{(\lambda^\prime_{p})}(\mathbf p^\prime_{p})
$
in which the four-momentum of the struck nucleon is defined through the off-shell
kinematic condition $p^\prime_p= p_d-p_n$, meaning that $p^{\prime 2}_p\ne m_N^2$
(see e.g.~\cite{Sargsian:2009hf}). This allows us to introduce the deuteron-wave-function
within VNA\cite{Sargsian:2009hf} as
\begin{equation}
	\Psi_d^{(\lambda_d;\lambda_1\lambda_2)}(\mathbf p)
		=
		-\frac{\bar{u}^{(\lambda_1)}(\mathbf p_1)\bar{u}^{(\lambda_2)}(\mathbf p_2)}{p_1^2-m_N^2}
		\frac{\Gamma_{dpn}\chi_d^{(\lambda_d)}(\mathbf p_d)}{\sqrt{2(2\pi)^32E_2}}
	\label{eqn:wfd}
\end{equation}
where
	$\mathbf p=\frac{1}{2}(\mathbf p_1 - \mathbf p_2)$
and
	$\mathbf p_d = \mathbf p_1 + \mathbf p_2$.
Further calculations we perform in the lab frame (i.e. the deuteron rest frame), for which
	$\mathbf p = \mathbf p_1 = -\mathbf p_2$.

Using Eq.~(\ref{eqn:wfd}) and 
introducing the invariant Feynman amplitude for vector meson photoproduction from the
struck nucleon, i.e.
\begin{equation}
	\mathcal{M}^{(\lambda_V\lambda_{p};\lambda_\gamma\lambda^\prime_{p})}_{\gamma N\rightarrow VN}
		(s_{\gamma N^*},t_{\gamma N^*})
		=
		\bar{u}^{(\lambda_{p})}(\mathbf p_{p})
		{\phi^\dagger}^{(\lambda_V)}_\nu(\mathbf p_V)
		\Gamma^{\mu\nu}_{\gamma N\rightarrow VN}
		\epsilon_\mu(q)
		u^{(\lambda^\prime_{p})}(\mathbf p^\prime_{p})
\end{equation}
into Eq.~(\ref{eqn:M0_a}), we find the PWIA amplitude to be
\begin{equation}
	\mathcal{M}^{(\lambda_{V}\lambda_{p}\lambda_{n};\lambda_\gamma\lambda_d)}_0
		=
		\sqrt{2(2\pi)^32E_{p}}
		\sum_{\lambda^\prime_p}
		\mathcal{M}^{(\lambda_V\lambda_{p};\lambda_\gamma\lambda^\prime_{p})}_{\gamma N\rightarrow VN}
			(s_{\gamma N^*},t_{\gamma N^*})
		\Psi_d^{(\lambda_d;\lambda^\prime_{p}\lambda_{n})}(\mathbf p_{n})
	\label{eqn:M0_b}
\end{equation}
where $s_{\gamma N^*}$ and $t_{\gamma N^*}$ are the Mandelstam variables at the
$\gamma N \rightarrow VN$ vertex.
Note that the vector meson photoproduction amplitude that enters above is half-off-shell
since the struck nucleon is initially in a virtual state. In these calculations, however,
we use on-shell spinors for the bound nucleon and account for off-shell effects only
kinematically by identifying the four-momentum of the bound nucleon as
$p^\prime_p=p_d-p_n$. Earlier estimates\cite{Sargsian:2009hf} demonstrated this approximation
to be reasonable when
$\frac{|\mathbf p^\prime_p|}{\sqrt{-t}},\frac{|\mathbf p^\prime_p|}{\sqrt{s}}\ll1$,
both of which are satisfied here.

\subsection{Single Rescattering Contribution}
\label{sec:singles}

\begin{figure}[hpt]
	\centering\includegraphics[scale=0.8]{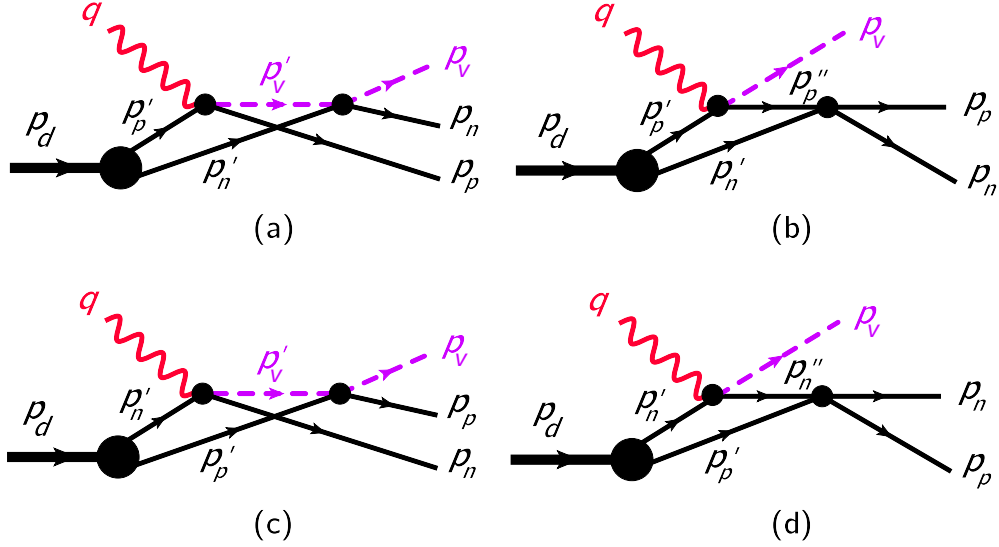}
	\caption{
		(Color online)
		Feynman diagrams for single rescattering contribution.
	}
	\label{fig:resc1}
\end{figure}

Within GEA, there are four single rescattering processes to consider
(Fig.~\ref{fig:resc1}). The processes can be separated into two groups. In the first group
(Fig.~\ref{fig:resc1}(a,b)), the proton receives its large momentum due to a hard
photoproduction vertex, while in the second (Fig.~\ref{fig:resc1}(c,d)) its large momentum
comes from a hard rescattering vertex.

While the first three diagrams in Fig.~\ref{fig:resc1} can be identified as having one
hard and one soft rescattering vertex, the fourth one requires two hard vertices in order
to provide a large momentum to both the vector meson and proton. In principle, a possible
``one-hard''and ``one-soft'' scenario can be realized for the fourth diagram if the high
momentum of the intermediate neutron is transferred to the final proton by means of a
small angle
charge interchange reaction. However, this contribution is negligible due to the pion
exchange nature of charge interchange scattering at the considered energies, which results
in an overall contribution that decreases with $s_{pn}$ and contributes at most a few
percent in the forward direction\footnote{
	A detailed discussion of charge interchange FSI for $d(e,e^\prime N)N$ is given
	in Ref.\cite{Sargsian:2009hf}.
}.

It is worth noting that due to antisymmetry of the nuclear wave function, the amplitude of
Fig.~\ref{fig:resc1}({c}) enters with the opposite sign relative to the amplitude of
Fig.~\ref{fig:resc1}({a}). However its contribution will be suppressed for near-threshold
production of heavy vector mesons (such as $J/\Psi$), since the high $-t_{\mathrm{thr.}}$
of the photoproduction reaction will make the first vertex hard. Thus we expect
appreciable contribution from Fig.~\ref{fig:resc1}({c}) only for higher photon energies,
or when the produced vector meson is light.

We will calculate only the contributions of diagrams Fig.~\ref{fig:resc1}({a,b,c}), for
which we use effective Feynman rules (see Ref.~\cite{Sargsian:2001ax}). Denoting the
invariant amplitudes of the processes as $\mathcal{M}_{1a}$, $\mathcal{M}_{1b}$ and
$\mathcal{M}_{1c}$, we obtain
\begin{align}
	\mathcal{M}_{1a}^{(\lambda_V,\lambda_p,\lambda_n;\lambda_\gamma,\lambda_d)}
		&=
		-\bar{u}^{(\lambda_p)}(\mathbf{p}_p)\bar{u}^{(\lambda_n)}(\mathbf{p}_n)\phi_\pi^{(\lambda_V)}(\mathbf{p}_V)
		\int\frac{d^4p_n^\prime}{(2\pi)^4i}\bigg[
		\Gamma^{\rho\pi}_{VN\rightarrow VN}
		\frac{G_{\nu\rho}(p_V^\prime)}{(p_V^\prime)^2-m_V^2+i\epsilon}
		\Gamma^{\mu\nu}_{\gamma N\rightarrow VN}
		\epsilon_\mu^{(\lambda_\gamma)}
	\notag \\ &\times
		\frac{\slashed{p}_p^\prime+m_N}{(p_p^\prime)^2-m_N^2+i\epsilon}
		\frac{\slashed{p}_n^\prime+m_N}{(p_n^\prime)^2-m_N^2+i\epsilon}
		\Gamma_{dpn}\chi_d^{(\lambda_d)}
		\bigg]
	\label{eqn:M1afull}
	\\
	\mathcal{M}_{1b}^{(\lambda_V,\lambda_p,\lambda_n;\lambda_\gamma,\lambda_d)}
		&=
		-\bar{u}^{(\lambda_p)}(\mathbf{p}_p)\bar{u}^{(\lambda_n)}(\mathbf{p}_n)\phi_\nu^{(\lambda_V)}(\mathbf{p}_V)
		\int\frac{d^4p_p^\prime}{(2\pi)^4i}\bigg[
		\Gamma_{pn\rightarrow pn}
		\frac{\slashed{p}_p^{\prime\prime}+m_N}{(p_p^{\prime\prime})^2-m_N^2+i\epsilon}
		\frac{\slashed{p}_n^\prime+m_N}{(p_n^\prime)^2-m_N^2+i\epsilon}
	\notag \\ &\times
		\Gamma^{\mu\nu}_{\gamma N\rightarrow VN}
		\epsilon_\mu^{(\lambda_\gamma)}
		\frac{\slashed{p}_p^\prime+m_N}{(p_p^\prime)^2-m_N^2+i\epsilon}
		\Gamma_{dpn}\chi_d^{(\lambda_d)}
		\bigg]
	\label{eqn:M1bfull}
	\\
	\mathcal{M}_{1c}^{(\lambda_V,\lambda_p,\lambda_n;\lambda_\gamma,\lambda_d)}
		&=
		-\bar{u}^{(\lambda_p)}(\mathbf{p}_p)\bar{u}^{(\lambda_n)}(\mathbf{p}_n)\phi_\pi^{(\lambda_V)}(\mathbf{p}_V)
		\int\frac{d^4p_p^\prime}{(2\pi)^4i}\bigg[
		\Gamma^{\rho\pi}_{VN\rightarrow VN}
		\frac{G_{\nu\rho}(p_V^\prime)}{(p_V^\prime)^2-m_V^2+i\epsilon}
		\Gamma^{\mu\nu}_{\gamma N\rightarrow VN}
		\epsilon_\mu^{(\lambda_\gamma)}
	\notag \\ &\times
		\frac{\slashed{p}_p^\prime+m_N}{(p_p^\prime)^2-m_N^2+i\epsilon}
		\frac{\slashed{p}_n^\prime+m_N}{(p_n^\prime)^2-m_N^2+i\epsilon}
		\Gamma_{dpn}\chi_d^{(\lambda_d)}
		\bigg]
	\label{eqn:M1cfull}
%	\\
%	\mathcal{M}_{1d}^{(\lambda_V,\lambda_p,\lambda_n;\lambda_\gamma,\lambda_d)}
%		&=
%		-\bar{u}^{(\lambda_p)}(\mathbf{p}_p)\bar{u}^{(\lambda_n)}(\mathbf{p}_n)\phi_\nu^{(\lambda_V)}(\mathbf{p}_V)
%		\int\frac{d^4p_p^\prime}{(2\pi)^4i}\bigg[
%		\Gamma_{pn\rightarrow pn}
%		\frac{\slashed{p}_p^\prime+m_N}{(p_p^\prime)^2-m_N^2+i\epsilon}
%		\frac{\slashed{p}_n^\prime+m_N}{(p_n^\prime)^2-m_N^2+i\epsilon}
%	\notag \\ &\times
%		\Gamma^{\mu\nu}_{\gamma N\rightarrow VN}
%		\epsilon_\mu^{(\lambda_\gamma)}
%		\frac{\slashed{p}_n^{\prime\prime}+m_N}{(p_n^{\prime\prime})^2-m_N^2+i\epsilon}
%		I_3(n)\Gamma_{dpn}\chi_d^{(\lambda_d)}
%		\bigg]
%	\label{eqn:M1dfull}
\end{align}
where $G_{\mu\nu}(p_V) = g_{\mu\nu}-(p_V)_\mu(p_V)_\nu/m_V^2$ is the propagator of the 
intermediate vector meson. The momentum notations of intermediate particles are given in
Fig.~\ref{fig:resc1}.

Our derivations follow the prescription of virtual nucleon approximation
(cf. Sec.~\ref{sec:PWIA} and Ref.~\cite{Sargsian:2009hf}), in which the spectator to the
photoproduction sub-reaction is placed on its mass shell by considering only the
positive-energy pole in the spectator energy integration, i.e. we take
\begin{equation}
	\int\frac{dp^{\prime 0}_{p/n}}{p^{\prime 2}_{p/n}-m_N^2+i\epsilon}
		= -i\frac{2\pi}{2E^\prime_{p/n}}
\end{equation}
where $E^\prime_{p/n} = +\sqrt{m_N^2+\mathbf{p}^{\prime 2}_{p/n}}$. Since this integration 
places the spectator on its mass shell, the completeness relation
$
	\slashed{p}^\prime_{p/n}+m_N
	=
	\sum_{\lambda^\prime_{p/n}}
	u^{(\lambda^\prime_{p/n})}(\mathbf{p}^\prime_{p/n})
	\bar{u}^{(\lambda^\prime_{p/n})}(\mathbf{p}^\prime_{p/n})
$
is exact. This, together with the approximate completeness relation used for the off-shell
nucleon propagator allows us to introduce the deuteron wave-function of
Eq.~(\ref{eqn:wfd}) into the rescattering amplitudes as follows:
\begin{align}
	\mathcal{M}_{1a}^{(\lambda_V,\lambda_p,\lambda_n;\lambda_\gamma,\lambda_d)}
		&=
		-\bar{u}^{(\lambda_p)}(\mathbf{p}_p)\bar{u}^{(\lambda_n)}(\mathbf{p}_n)\phi_\pi^{(\lambda_V)}(\mathbf{p}_V)
		\sum_{\lambda_p^\prime,\lambda_n^\prime}
		\int\frac{d^3\mathbf{p}_n^\prime}{(2\pi)^3}\bigg[
		\sqrt{\frac{2(2\pi)^3}{2E_n^\prime}}
		\Gamma^{\rho\pi}_{VN\rightarrow VN}
		u^{(\lambda_n^\prime)}(\mathbf{p}_n^\prime)
	\notag \\ &\times
		\frac{G_{\nu\rho}(p_V^\prime)}{(p_V^\prime)^2-m_V^2+i\epsilon}
		\Gamma^{\mu\nu}_{\gamma N\rightarrow VN}
		\epsilon_\mu^{(\lambda_\gamma)}
		u^{(\lambda_p^\prime)}(\mathbf{p}_p^\prime)
		\Psi_d^{(\lambda_p^\prime,\lambda_n^\prime;\lambda_d)}(\mathbf{p}_n^\prime)
		\bigg]
	\label{eqn:M1adwf}
	\\
	\mathcal{M}_{1b}^{(\lambda_V,\lambda_p,\lambda_n;\lambda_\gamma,\lambda_d)}
		&=
		-\bar{u}^{(\lambda_p)}(\mathbf{p}_p)\bar{u}^{(\lambda_n)}(\mathbf{p}_n)\phi_\nu^{(\lambda_V)}(\mathbf{p}_V)
		\int\frac{d^3\mathbf{p}_n^\prime}{(2\pi)^3}\bigg[
		\sqrt{\frac{2(2\pi)^3}{2E_n^\prime}}
		\Gamma_{pn\rightarrow pn}
		u^{(\lambda_n^\prime)}(\mathbf{p}_n^\prime)
	\notag \\ &\times
		\frac{\slashed{p}_p^{\prime\prime}+m_N}{(p_p^{\prime\prime})^2-m_N^2+i\epsilon}
		\Gamma^{\mu\nu}_{\gamma N\rightarrow VN}
		\epsilon_\mu^{(\lambda_\gamma)}
		u^{(\lambda_p^\prime)}(\mathbf{p}_p^\prime)
		\Psi_d^{(\lambda_p^\prime,\lambda_n^\prime;\lambda_d)}(\mathbf{p}_n^\prime)
		\bigg]
	\label{eqn:M1bdwf}
	\\
	\mathcal{M}_{1c}^{(\lambda_V,\lambda_p,\lambda_n;\lambda_\gamma,\lambda_d)}
		&=
		-\bar{u}^{(\lambda_p)}(\mathbf{p}_p)\bar{u}^{(\lambda_n)}(\mathbf{p}_n)\phi_\pi^{(\lambda_V)}(\mathbf{p}_V)
		\int\frac{d^3\mathbf{p}_p^\prime}{(2\pi)^3}\bigg[
		\sqrt{\frac{2(2\pi)^3}{2E_p^\prime}}
		\Gamma^{\rho\pi}_{VN\rightarrow VN}
		u^{(\lambda_p^\prime)}(\mathbf{p}_p^\prime)
	\notag \\ &\times
		\frac{G_{\nu\rho}(p_V^\prime)}{(p_V^\prime)^2-m_V^2+i\epsilon}
		\Gamma^{\mu\nu}_{\gamma N\rightarrow VN}
		\epsilon_\mu^{(\lambda_\gamma)}
		u^{(\lambda_n^\prime)}(\mathbf{p}_n^\prime)
		\Psi_d^{(\lambda_n^\prime,\lambda_p^\prime;\lambda_d)}(\mathbf{p}_p^\prime)
		\bigg].
	\label{eqn:M1cdwf}
\end{align}
Note
that, due to the fact the deuteron  wave-function is antisymmetric with respect to a $p$-$n$
permutation, the amplitude $\mathcal{M}_{1c}$ has a negative sign relative to
$\mathcal{M}_{1a}$ and $\mathcal{M}_{1b}$.

We consider next the remaining ``fast'' propagator in each amplitude: the vector meson
propagators for $\mathcal{M}_{1a}$ and $\mathcal{M}_{1c}$ and the proton propagator for
$\mathcal{M}_{1b}$.
In each case we introduce the momentum transfer at the rescattering vertex as
$K = p_n - p_n^\prime$ and rewrite the propagators in terms of it.

For the propagator in $\mathcal{M}_{1a}$, we obtain
\begin{align}
	(p_V^\prime)^2-m_V^2+i\epsilon
		&= (p_V+p_n-p_n^\prime)^2 - m_V^2 + i\epsilon
		= (p_V+K)^2-m_V^2+i\epsilon
	\notag \\
		&= 2p_{V,z}\left(\Delta_{1a}-K_z+i\epsilon\right)
	\label{eqn:resc1ka} ,
\end{align}
where
\begin{equation}
	\Delta_{1a} = \frac{K^2+2K_0E_V-2\mathbf{K}_\perp\cdot\mathbf{p}_V}{2p_{V,z}} .
	\label{eqn:delta1a}
\end{equation}
To proceed with the integration, we use the fact that the rescattering amplitude is
dominated by small angle scattering. This means that $K_z^2 \ll \mathbf{K}_\perp^2$,
so $K_0$ and $K^2$ are approximately equal to their $K_z=0$ values. Moreover, 
$\frac{\partial\Delta_{1a}}{\partial K_z}\sim\frac{K_z}{p_{V,z}}\ll1$
so we may treat $\Delta_{1a}$ as independent of $K_z$, effectively linearizing
the denominator of Eq.~(\ref{eqn:resc1ka}). With this in mind, and the fact that the
integration over $\mathbf{p}_n^\prime$ in Eq.~(\ref{eqn:M1adwf}) can be rewritten as
an integration over $\mathbf{K}$, the $dK_z$ integration can be performed using
\begin{equation}
	\int\frac{f(z)dz}{\Delta-z+i\epsilon}
	=
	-i\pi f(\Delta)
	+
	\mathcal P\int\frac{f(z)dz}{\Delta-z}
	\label{eqn:decomp} ,
\end{equation}
where the symbol $\mathcal P$ indicates that the Cauchy principal value of the integral is
to be taken.

In applying the decomposition (\ref{eqn:decomp}) to $\mathcal{M}_{1a}$, we can separate
$\mathcal{M}_{1a}$ into on-shell and off-shell parts, since the condition
$\Delta_{1a}=K_z$ imposed by the delta function corresponds to the on-shell condition for
the intermediate vector meson---i.e. it happens when $(p_V^\prime)^2=m_V^2$. We shall
henceforth refer to this part of $\mathcal{M}_{1a}$ as the pole term. Because, in this
term, the internal vector meson line is on its mass shell, we can now use the completeness
relation
$
	G_{\nu\rho}(p_V^\prime)
	= \sum_{\lambda_V^\prime}
	{\phi_\nu}^{(\lambda_V^\prime)}(\mathbf p_V^\prime)
	{\phi_\rho^*}^{(\lambda_V^\prime)}(\mathbf p_V^\prime)
$
and gather terms together into the invariant amplitudes for the $\gamma N\rightarrow VN$
and $VN\rightarrow VN$ transitions:
\begin{align}
	\mathcal{M}_{1a,\mathrm{pole}}^{(\lambda_V,\lambda_p,\lambda_n;\lambda_\gamma,\lambda_d)}
		&=
		\frac{i}{4p_{V,z}}
		\sum_{\lambda_V^\prime,\lambda_p^\prime,\lambda_n^\prime}
		\int\frac{d^2\mathbf{K}_\perp}{(2\pi)^2}\bigg[
		\sqrt{\frac{2(2\pi)^3}{2E_n^\prime}}
		\mathcal{M}_{VN\rightarrow VN}^{(\lambda_V,\lambda_n;\lambda_V^\prime,\lambda_n^\prime)}(s_{VN},t_{VN})
	\notag \\ &\times
		\mathcal{M}_{\gamma N\rightarrow VN}^{(\lambda_V^\prime,\lambda_p;\lambda_\gamma,\lambda_p^\prime)}
			(s_{\gamma N^*},t_{\gamma N^*})
		\Psi_d^{(\lambda_p^\prime,\lambda_n^\prime;\lambda_d)}(p_{n,z}-\Delta_{1a},\mathbf{p}_{n,\perp}-\mathbf{K}_\perp)
		\bigg]
	.
	\label{eqn:M1apole}
\end{align}
Here, as in PWIA the invariant amplitudes appearing on the RHS are functions of the
Mandelstam variables for their respective transitions.

For the other two rescattering amplitudes, similar decompositions of the fast propagators
are possible. For $\mathcal{M}_{1b}$, we have
\begin{align}
	(p_p^\prime)^2-m_N^2+i\epsilon
		&= (p_p+p_n-p_n^\prime)^2-m_N^2+i\epsilon
		= (p_p+K)^2-m_N^2+i\epsilon
		\notag \\ &= 2p_{p,z}\left(\Delta_{1b}-K_z+i\epsilon\right)
	\label{eqn:resc1kb} ,
\end{align}
where
\begin{equation}
	\Delta_{1b} = \frac{K^2+2K_0E_p-2\mathbf{K}_\perp\cdot\mathbf{p}_p}{2p_{p,z}} .
	\label{eqn:delta1b}
\end{equation}
This gives us a pole part of the amplitude equal to
\begin{align}
	\mathcal{M}_{1b,\mathrm{pole}}^{(\lambda_V,\lambda_p,\lambda_n;\lambda_\gamma,\lambda_d)}
		&=
		\frac{i}{4p_{p,z}}
		\sum_{\lambda_p^\prime,\lambda_p^{\prime\prime},\lambda_n^\prime}
		\int\frac{d^2\mathbf{K}_\perp}{(2\pi)^2}\bigg[
		\sqrt{\frac{2(2\pi)^3}{2E_n^\prime}}
		\mathcal{M}_{pn\rightarrow pn}^{(\lambda_p,\lambda_n;\lambda_p^{\prime\prime},\lambda_n^\prime)}(s_{pn},t_{pn})
	\notag \\ &\times
		\mathcal{M}_{\gamma N\rightarrow VN}^{(\lambda_V,\lambda_p^{\prime\prime};\lambda_\gamma,\lambda_p^\prime)}
			(s_{\gamma N^*},t_{\gamma N^*})
		\Psi_d^{(\lambda_p^\prime,\lambda_n^\prime;\lambda_d)}(p_{n,z}-\Delta_{1b},\mathbf{p}_{n,\perp}-\mathbf{K}_\perp)
		\bigg]
	.
	\label{eqn:M1bpole}
\end{align}
As for $\mathcal{M}_{1c}$, we have
\begin{align}
	(p_V^\prime)^2-m_V^2+i\epsilon
		&= (q+p_n^\prime-p_n)^2-m_V^2+i\epsilon
		= (q-K)^2-m_V^2+i\epsilon
		\notag \\ &= 2q_0\left(K_z-\Delta_{1c}+i\epsilon\right)
	\label{eqn:resc1kc}
\end{align} ,
where
\begin{equation}
	\Delta_{1c} = \frac{m_V^2-K^2}{2q_0}+K_0,
	\label{eqn:delta1c}
\end{equation}
which gives us a pole part of the amplitude equal to
\begin{align}
	\mathcal{M}_{1c,\mathrm{pole}}^{(\lambda_V,\lambda_p,\lambda_n;\lambda_\gamma,\lambda_d)}
		&=
		\frac{i}{4q_0}
		\sum_{\lambda_V^\prime,\lambda_p^\prime,\lambda_n^\prime}
		\int\frac{d^2\mathbf{K}_\perp}{(2\pi)^2}\bigg[
		\sqrt{\frac{2(2\pi)^3}{2E_p^\prime}}
		\mathcal{M}_{VN\rightarrow VN}^{(\lambda_V,\lambda_p;\lambda_V^\prime,\lambda_p^\prime)}(s_{VN},t_{VN})
	\notag \\ &\times
		\mathcal{M}_{\gamma N\rightarrow VN}^{(\lambda_V^\prime,\lambda_n;\lambda_\gamma,\lambda_n^\prime)}
			(s_{\gamma N^*},t_{\gamma N^*})
		\Psi_d^{(\lambda_n^\prime,\lambda_p^\prime;\lambda_d)}(-p_{n,z}+\Delta_{1c},-\mathbf{p}_{n,\perp}+\mathbf{K}_\perp)
		\bigg]
	.
	\label{eqn:M1cpole}
\end{align}
For the principal value (PV) parts of each amplitude, we simplify the notation by
introducing half-off-shell $VN\rightarrow VN$ and $pn\rightarrow pn$ amplitudes that allow us to write the following:
\begin{align}
	\mathcal{M}_{1a,\mathrm{pv}}^{(\lambda_V,\lambda_p,\lambda_n;\lambda_\gamma,\lambda_d)}
		&=
		\frac{1}{2p_{V,z}}
		\sum_{\lambda_V^\prime,\lambda_p^\prime,\lambda_n^\prime}
		\int\frac{d^2\mathbf{K}_\perp}{(2\pi)^2}\mathcal{P}\int\frac{dK_z}{2\pi}\bigg[
		\sqrt{\frac{2(2\pi)^3}{2E_n^\prime}}
		\mathcal{M}_{V^*N\rightarrow VN}^{(\lambda_V\lambda_n;\lambda_V^\prime\lambda_n^\prime)}(s_{V^*N},t_{V^*N})
	\notag \\ &\times
		\mathcal{M}_{\gamma N^*\rightarrow VN}^{(\lambda_V^\prime\lambda_p;\lambda_\gamma\lambda_p^\prime)}
			(s_{\gamma N^*},t_{\gamma N^*})
		\frac{\Psi_d^{(\lambda_p^\prime,\lambda_n^\prime;\lambda_d)}(p_{n,z}-\Delta_{1a},\mathbf{p}_{n,\perp}-\mathbf{K}_\perp)}
			{K_z - \Delta_{1a}}
		\bigg]
	\label{eqn:M1apv}
	\\
	\mathcal{M}_{1b,\mathrm{pv}}^{(\lambda_V,\lambda_p,\lambda_n;\lambda_\gamma,\lambda_d)}
		&=
		\frac{1}{2p_{p,z}}
		\sum_{\lambda_V^\prime,\lambda_p^\prime,\lambda_n^\prime}
		\int\frac{d^2\mathbf{K}_\perp}{(2\pi)^2}\mathcal{P}\int\frac{dK_z}{2\pi}\bigg[
		\sqrt{\frac{2(2\pi)^3}{2E_n^\prime}}
		\mathcal{M}_{p^*n\rightarrow pn}^{(\lambda_p\lambda_n;\lambda_p^{\prime\prime}\lambda_n^\prime)}(s_{p^*n},t_{p^*n})
	\notag \\ &\times
		\mathcal{M}_{\gamma N^*\rightarrow VN}^{(\lambda_V\lambda_{p^\prime\prime};\lambda_\gamma\lambda_p^\prime)}
			(s_{\gamma N^*},t_{\gamma N^*})
		\frac{\Psi_d^{(\lambda_p^\prime,\lambda_n^\prime;\lambda_d)}(p_{n,z}-\Delta_{1b},-\mathbf{p}_{n,\perp}+\mathbf{K}_\perp)}
			{K_z - \Delta_{1b}}
		\bigg]
	\label{eqn:M1bpv}
	\\
	\mathcal{M}_{1c,\mathrm{pv}}^{(\lambda_V,\lambda_p,\lambda_n;\lambda_\gamma,\lambda_d)}
		&=
		-\frac{1}{2q_0}
		\sum_{\lambda_V^\prime,\lambda_p^\prime,\lambda_n^\prime}
		\int\frac{d^2\mathbf{K}_\perp}{(2\pi)^2}\mathcal{P}\int\frac{dK_z}{2\pi}\bigg[
		\sqrt{\frac{2(2\pi)^3}{2E_p^\prime}}
		\mathcal{M}_{V^*N\rightarrow VN}^{(\lambda_V\lambda_p;\lambda_V^\prime\lambda_p^\prime)}(s_{V^*N},t_{V^*N})
	\notag \\ &\times
		\mathcal{M}_{\gamma N^*\rightarrow VN}^{(\lambda_V^\prime\lambda_n;\lambda_\gamma\lambda_n^\prime)}
			(s_{\gamma N^*},t_{\gamma N^*})
		\frac{\Psi_d^{(\lambda_n^\prime,\lambda_p^\prime;\lambda_d)}(-p_{n,z}+\Delta_{1c},\mathbf{p}_{n,\perp}-\mathbf{K}_\perp)}
			{K_z - \Delta_{1c}}
		\bigg]
	\label{eqn:M1cpv} ,
\end{align}
where we have written $V^*$ and $p^*$ to indicate that these particles are off-shell in
their intermediate states. We will estimate these amplitudes numerically by substituting
off-shell rescattering amplitudes with on-shell counterparts evaluated at off-shell
kinematics. However, it is important to note that our approach can only be used to probe
the $VN\rightarrow VN$ amplitude when the principal value parts are negligible.

\subsection{Double Rescattering Contribution}
\label{sec:doubles}

\begin{figure}[hpt]
	\centering\includegraphics[scale=0.8]{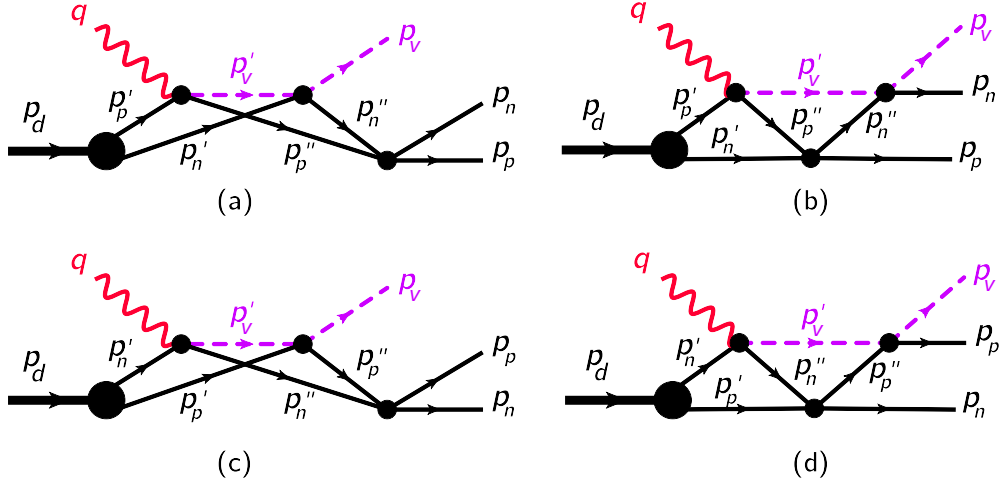}
	\caption{
		(Color online)
		Feynman diagrams for double rescattering contribution.
	}
	\label{fig:resc2}
\end{figure}

The last term in Eq.~(\ref{M012}) corresponds to double rescattering contributions
(Fig.~\ref{fig:resc2}). They, like the single rescattering processes, can be split into
two groups. The first group (Fig.~\ref{fig:resc2}(a,b)) correspond to hard vector meson
production from the proton, followed by two consecutive soft rescatterings from the slow
neutron. The two remaining diagrams (Fig.~\ref{fig:resc2}(c,d)) correspond to vector meson
production from the neutron, and provide the proton with a large final momentum by means
of a hard rescattering vertex.

It is worth noting that, for heavy vector meson production near threshold, the last two
diagrams are negligible due to the large $-t$ involved in both the photoproduction vertex
and one of the proton rescatterings. However, the contribution of 
Fig.~\ref{fig:resc2}({c}) becomes relevant for high energies. The contribution of
Fig.~\ref{fig:resc2}({d}) remains negligible at high energies due to the fact that it
contains an intermediate state with a slow proton-neutron pair, the momentum of which is
integrated over. This warrants the application of a closure condition as an approximation,
which results in a cancelation of any correction this diagram would make to the
contribution of
Fig.~\ref{fig:resc1}(b)\footnote{
	This is the same approximation by which the interactions between scatterers in the
	target are neglected in the conventional Glauber theory\cite{Frankfurt:1995rq} or in
	GEA\cite{Sargsian:2001ax}.
}.

In calculating the diagrams Fig.~\ref{fig:resc2}(a,b,c), we will concentrate only on the
contributions from the pole terms of the fast hadron propagators that contain on-shell
$p$-$n$ and $V$-$N$ scattering amplitudes. As discussed in the previous section, this is
justified by focusing only on kinematics where the principal value contributions are a
small correction. Since one would expect that $\mathcal{M}_2\ll \mathcal{M}_1$, the
PV parts of $\mathcal{M}_2$ should be significantly smaller than the overall amplitude.

Applying the effective Feynman rules and performing the loop integrations in accordance
with the same prescription used for the single rescattering amplitudes, we find the pole
terms of the diagrams in Fig.~\ref{fig:resc2}(a,b,c) to be
\begin{align}
	\mathcal{M}_{2a}^{(\lambda_V,\lambda_p,\lambda_n;\lambda_\gamma,\lambda_d)}
		&=
		-\frac{1}{4p_{p,z}p_{V,z}}
		\sum_{\mathrm{spins}}
		\int\frac{d^2\mathbf{K}}{(2\pi)^2}
		\int\frac{d^2\mathbf{K}^\prime}{(2\pi)^2}
		\bigg[
		\sqrt{\frac{2(2\pi)^3}{2E_n^\prime}}
		\frac{1}{2E_n^{\prime\prime}}
		\mathcal{M}_{pn\rightarrow pn}^{(\lambda_p,\lambda_n;\lambda_p^{\prime\prime},\lambda_n^{\prime\prime})}
			(p_p,p_n;p_p^{\prime\prime},p_n^{\prime\prime})
	\notag \\ &\times
		\mathcal{M}_{VN\rightarrow VN}^{(\lambda_V,\lambda_n^{\prime\prime};\lambda_V^\prime,\lambda_n^\prime)}
			(p_V,p_n^{\prime\prime};p_V^\prime,p_n^\prime)
		\mathcal{M}_{\gamma N\rightarrow VN}^{(\lambda_V^\prime,\lambda_p^{\prime\prime};\lambda_\gamma,\lambda_p^\prime)}
			(p_V^\prime,p_p^{\prime\prime};p_\gamma,p_p^\prime)
	\notag \\ &\times
		\Psi_d^{(\lambda_p^\prime,\lambda_n^\prime;\lambda_d)}
			(p_{n,z}-\Delta_{2a}-\Delta_{2a}^\prime;\mathbf{p}_{n,\perp}-\mathbf{K}_\perp-\mathbf{K}^\prime_\perp)
		\bigg]
	\label{eqn:M2a}
	\\
	\mathcal{M}_{2b}^{(\lambda_V,\lambda_p,\lambda_n;\lambda_\gamma,\lambda_d)}
		&=
		-\frac{1}{4p_{p,z}p_{V,z}}
		\sum_{\mathrm{spins}}
		\int\frac{d^2\mathbf{K}}{(2\pi)^2}
		\int\frac{d^2\mathbf{K}^\prime}{(2\pi)^2}
		\bigg[
		\sqrt{\frac{2(2\pi)^3}{2E_n^\prime}}
		\frac{1}{2E_n^{\prime\prime}}
		\mathcal{M}_{VN\rightarrow VN}^{(\lambda_V,\lambda_p;\lambda_V^\prime,\lambda_p^{\prime\prime})}
			(p_V,p_p;p_V^\prime,p_p^{\prime\prime})
	\notag \\ &\times
		\mathcal{M}_{pn\rightarrow pn}^{(\lambda_p,\lambda_n^{\prime\prime};\lambda_p^{\prime\prime},\lambda_n^\prime)}
			(p_p,p_n^{\prime\prime};p_p^{\prime\prime},p_n^\prime)
		\mathcal{M}_{\gamma N\rightarrow VN}^{(\lambda_V^\prime,\lambda_p^{\prime\prime};\lambda_\gamma,\lambda_p^\prime)}
			(p_V^\prime,p_p^{\prime\prime};p_\gamma,p_p^\prime)
	\notag \\ &\times
		\Psi_d^{(\lambda_p^\prime,\lambda_n^\prime;\lambda_d)}
			(p_{n,z}-\Delta_{2b}-\Delta_{2b}^\prime;\mathbf{p}_{n,\perp}-\mathbf{K}_\perp-\mathbf{K}^\prime_\perp)
		\bigg]
	\label{eqn:M2b}
	\\
	\mathcal{M}_{2c}^{(\lambda_V,\lambda_p,\lambda_n;\lambda_\gamma,\lambda_d)}
		&=
		-\frac{1}{4q_0p_{p,z}}
		\sum_{\mathrm{spins}}
		\int\frac{d^2\mathbf{K}}{(2\pi)^2}
		\int\frac{d^2\mathbf{K}^\prime}{(2\pi)^2}
		\bigg[
		\sqrt{\frac{2(2\pi)^3}{2E_p^\prime}}
		\frac{1}{2E_n^{\prime\prime}}
		\mathcal{M}_{pn\rightarrow pn}^{(\lambda_p,\lambda_n;\lambda_p^{\prime\prime},\lambda_n^{\prime\prime})}
			(p_p,p_n;p_p^{\prime\prime},p_n^{\prime\prime})
	\notag \\ &\times
		\mathcal{M}_{VN\rightarrow VN}^{(\lambda_V,\lambda_p^{\prime\prime};\lambda_V^\prime,\lambda_p^\prime)}
			(p_V,p_p^{\prime\prime};p_V^\prime,p_p^\prime)
		\mathcal{M}_{\gamma N\rightarrow VN}^{(\lambda_V^\prime,\lambda_p^{\prime\prime};\lambda_\gamma,\lambda_p^\prime)}
			(p_V^\prime,p_n^{\prime\prime};p_\gamma,p_n^\prime)
	\notag \\ &\times
		\Psi_d^{(\lambda_n^\prime,\lambda_p^\prime;\lambda_d)}
			(-p_{n,z}+\Delta_{2a}+\Delta_{2a}^\prime;-\mathbf{p}_{n,\perp}+\mathbf{K}_\perp+\mathbf{K}^\prime_\perp)
		\bigg]
	\label{eqn:M2c}
\end{align}
where the delta factors are
\begin{align}
	\Delta_{2a} &= \frac{K^2+2E_pK_0-2\mathbf{p}_p\cdot\mathbf{K}_\perp}{2p_{p,z}} \\
	\Delta_{2a}^\prime &= \frac{(K^\prime)^2+2E_VK^\prime_0-2\mathbf{p}_V\cdot\mathbf{K}^\prime_\perp}{2p_{V,z}} \\
	\Delta_{2b} &= \frac{K^2+2E_VK_0-2\mathbf{p}_V\cdot\mathbf{K}_\perp}{2p_{V,z}} \\
	\Delta_{2b}^\prime &= \frac{(K^\prime)^2+2E_pK^\prime_0-2\mathbf{p}_p\cdot\mathbf{K}^\prime_\perp}{2p_{p,z}} \\
	\Delta_{2c} &= \frac{K^2+2E_pK_0-2\mathbf{p}_p\cdot\mathbf{K}_\perp}{2p_{p,z}} \\
	\Delta_{2c}^\prime &= \frac{m_V^2-(K^\prime)^2}{2q_0} + K^\prime_0
\end{align}
A derivation of these results can be found in the appendix.

\subsection{Differential Cross Section}	
\label{sec:diffcrx}

The differential cross section of reaction (\ref{eqn:reaction}) is given by
Eq.~(\ref{eqn:crx}). We consider the particular case in which the vector meson and proton
are detected, and the neutron is inferred from missing mass. In accordance with this, we
integrate out the neutron momentum to obtain
\begin{equation}
	d\sigma
	=
	\frac{\overline{|\mathcal{M}|^2}}{4\sqrt{s_d-M_d^2}}
	(2\pi)\delta((p_d+q-p_V-p_{p})^2 - m_N^2)
	\frac{d^3\mathbf p_V}{(2E_V)(2\pi)^3}
	\frac{d^3\mathbf p_{p}}{(2E_{p})(2\pi)^3}
	.
\end{equation}
We eliminate the remaining delta function by integrating over the magnitude of the proton
momentum, which results in the following form of the differential cross-section:
\begin{eqnarray}
	\frac{d^5\sigma}{d^3\mathbf p_Vd\Omega_{p}}
	&=&
	\frac{\overline{|\mathcal{M}|^2}}{4\sqrt{s_d-M_d^2}}
	\frac{1}{(2\pi)^5}
	\frac{1}{8E_VE_{1f}E_{n}}
	\frac{p_{p}^3}{|\mathbf p_{p}\cdot(\mathbf v_{p}-\mathbf v_{n})|}
	.
	\label{eqn:diffcrx}
\end{eqnarray}
Here the outgoing nucleon velocities are $\mathbf v = \frac{\mathbf p}{E}$ and the total 
scattering amplitude is defined according to Eq.~(\ref{M012}).
 
\section{Numerical Estimates}
\label{sec:num}

For numerical estimates, it is convenient to express the invariant Feynman amplitude
$\mathcal{M}_{AB\rightarrow CD}^{(\lambda_C,\lambda_D;\lambda_A,\lambda_B)}$
for the process $AB\rightarrow CD$ through the diffractive amplitude
$f_{AB\rightarrow CD}^{(\lambda_C,\lambda_D;\lambda_A,\lambda_B)}$
in the following form:
\begin{equation}
	\mathcal{M}_{AB\rightarrow X}^{(\lambda_C,\lambda_D;\lambda_A,\lambda_B)}(s,t)
		= \sqrt{(s-(m_A-m_B)^2)(s-(m_A+m_B)^2)}
		f_{AB\rightarrow CD}^{(\lambda_C,\lambda_D;\lambda_A,\lambda_B)}(s,t)
	\label{amples} .
\end{equation}
Furthermore, we will parameterize the diffractive amplitude in the form:
\begin{equation}
	f_{AB\rightarrow CD}^{(\lambda_C,\lambda_D;\lambda_A,\lambda_B)}
	= A(s)(i+\alpha(s))e^{\frac{B(s)}{2}t}
	\delta^{\lambda_A\lambda_C}\delta^{\lambda_B\lambda_D}
	\label{eqn:diffamp}
	,
\end{equation}
which assumes helicity conservation, and which is most appropriate for the case of small
angle scattering. In the case that the reaction is elastic, $f$ is normalized so that
$\mathrm{Im} f(s,0) = \sigma_{\mathrm{tot}}(s)$, meaning in these cases
$A(s)=\sigma_{\mathrm{tot}}(s)$. Since helicity is conserved, we will hereafter omit
indices.

For the $pn\rightarrow pn$ amplitude, $\sigma_{\mathrm{tot}}$, $\alpha$, and $B$ are
determined for low energies using SAID data\cite{Arndt:2000xc} and for high energies using
a parameterization based on particle data\cite{Beringer:1900zz}.

For the $\gamma N \rightarrow VN$ amplitude, we use existing parameterizations both in the
invariant Feynman amplitude form or in the form of Eq.~(\ref{amples}).

Since the main goal of the present work is to investigate the feasibility of using
reaction (\ref{eqn:reaction}) to extract the amplitude of $VN\rightarrow VN$ scattering,
our main focus in this section will be to study the sensitivity of the cross section to
the parameters $\sigma_{\mathrm{tot}}^{VN}$ and $B_{VN}$. 

\subsection{Geometry of the reaction}
\label{sec:geometry}

\begin{figure}[hpt]
	\centering
	\includegraphics[scale=0.5]{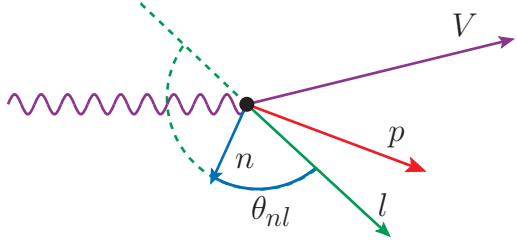}
	\caption{
		(Color online)
		Geometry of reaction (\ref{eqn:reaction}).
	}
	\label{fig:geometry}
\end{figure}

For simplicity, we consider the coplanar case, in which the three final
state particles are all contained in the same plane. This fixes two of the five
final-state degrees of freedom. The last three are the momentum transfer $t$, the final
neutron momentum $p_n$, and the angle $\theta_{nl}$ between the neutron direction and the
three-momentum transfer $\mathbf{l}$ (c.f.~Fig.~\ref{fig:geometry}).
When performing numerical estimates, we hold $t$
and $p_n$ fixed, and vary the value of $\theta_{nl}$ from zero to $180^\circ$, or at least
to whatever limit is kinematically allowed.

It is worth noting that $\mathbf{l}$ and $\mathbf{p}_p$ are not fixed, but are themselves
functions of $\theta_{nl}$. Both the direction and magnitude of $\mathbf{l}$ change, and
since $\mathbf{l} = \mathbf{p}_p + \mathbf{p}_n$, as $\theta_{nl}$ gets larger, so does
$\theta_{pl}$.

One expects rescattering peaks to occur where the fast scattered hadron is roughly
transverse with the scatterer. As can be seen in Fig.~\ref{fig:geometry}, this will occur
for fairly small $\theta_{nl}$ for the $V$-$n$ rescattering peak, but for larger
$\theta_{nl}$ for the $p$-$n$ rescattering peak. Thus, we expect as a general rule,
$\theta_{nl}^{pn} > \theta_{nl}^{VN}$, where the superscript indicates that this angle
is where a rescattering peak occurs.

\subsection{Photoproduction of $\phi$ mesons}
\label{sec:phi}

Recent studies of $\phi$ meson photoproduction from nuclei have generated much interest due to
a large $\phi$-$N$ cross-section observed  at near-threshold
energies\cite{Ishikawa:2004id,Wood:2010ei}. The $\phi$-$N$ cross-sections are usually
extracted by studying the $A$ dependence of the differential cross-section and employing
a Glauber-like analysis. The extracted results vary from $20$ to
$70~\mathrm{mb}$\cite{Qian:2010rr}, significantly exceeding the vector meson dominance
prediction of $10~\mathrm{mb}$.

\begin{figure}[hpt]
	\centering
	\centering\includegraphics[scale=0.5]{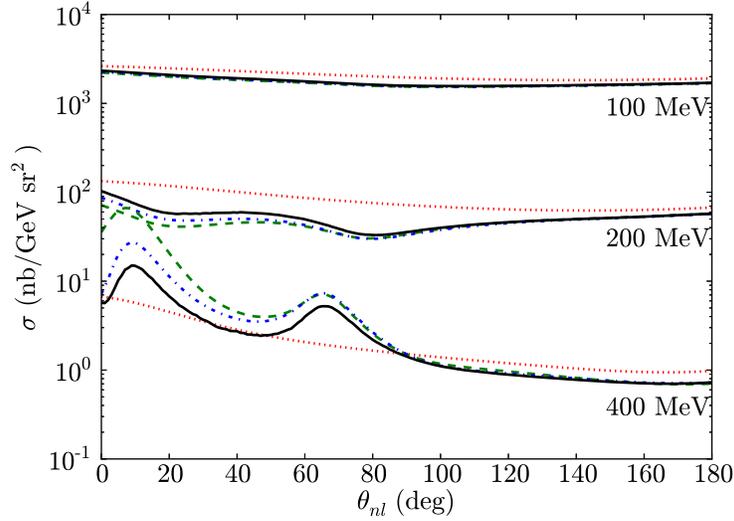}
	\caption{
		(Color online)
		Angular dependence of the $\phi$ photoproduction cross-section at different
		neutron momenta, assuming  $\sigma_{\phi N}=30~\mathrm{mb}$ and
		$B=10~\mathrm{GeV}^{-2}$.
		The dotted (red) line indicates PWIA. Dashed (green) indicates PWIA and the first
		two single rescattering terms (off-neutron production is neglected). Dash-dotted
		(blue) indicates PWIA + all single rescattering. Solid (black) indicates PWIA,
		single, and double rescattering.
		$q_0=5~\mathrm{GeV}$ and $-t=1.2~\mathrm{GeV}^2$.
	}
	\label{fig:phicrx}
\end{figure}

A deuteron target was used in several recent
experiments\cite{Mibe:2007aa,Qian:2009ab,Qian:2010rr}. In Ref.~\cite{Qian:2009ab}, the
analysis of the $d(\gamma,pK^+K^-)n$ exclusive reaction for photon energies in the range
$1.65-3.59~\mathrm{GeV}$ yielded a $\phi$-$N$ cross-section above $20~\mathrm{mb}$.
While measurements of the coherent reaction $d(\gamma,\phi)d$ clearly indicate the
presence of $\phi$-$N$ rescattering, they could not distinguish between the vector meson
dominance value of $\sigma_{\phi N}=10~\mathrm{mb}$ and the recently measured
$\sigma_{\phi N}=30~\mathrm{mb}$, provided the slope factor is assumed to be around
$10~\mathrm{GeV}^{-2}$\cite{Mibe:2007aa}.

To demonstrate the sensitivity of the break-up reaction to the $\phi$-$N$ scattering
parameters, we performed numerical evaluations of the cross-section (\ref{eqn:diffcrx}) at
energies for which a coherent $\phi$ production experiment was performed at
JLab\cite{Mibe:2007aa}.

In Fig.~\ref{fig:phicrx}, we present the angular distribution of the cross-section at
different values of the recoil neutron momentum, calculated assuming
$\sigma_{\phi N}=30~\mathrm{mb}$ and $B=10~\mathrm{GeV}^{-2}$. The cross-section value of
$30~\mathrm{mb}$ is suggested by recent experiments
(cf.~\cite{Ishikawa:2004id,Wood:2010ei}) and the value of the slope factor by an
analysis in Ref.~\cite{Mibe:2007aa}. Here we use the shorthand notation
\begin{equation}
	\sigma \equiv \frac{d^5\sigma}{dp_Vd\Omega_Vd\Omega_p}
	\label{eqn:sigma} .
\end{equation}

The figure shows an angular distribution qualitatively similar to what is observed in 
$d(e,e^\prime p)n$ reactions\cite{Sargsian:2009hf,Egiyan:2007qj,Boeglin:2011mt}: for
neutron momenta $\le 200~\mathrm{MeV}$ screening effects are dominant, as FSIs decrease
the overall cross-section due to destructive interference between the PWIA amplitude
$\mathcal{M}_0$ and the single rescattering amplitude
$\mathcal{M}_1 = \mathcal{M}_{1a} + \mathcal{M}_{1b}$.
For $p_n\ge 400~\mathrm{MeV}$ the cross-section is dominated by the square of single
rescattering amplitude, which results in an increase in the cross-section as compared to 
the PWIA value. The present case differs from the
$d(e,e^\prime p)n$ reaction, however, in that two
distinct screening minima are observed for $p_n=200~\mathrm{MeV}$, due to the presence of
both a $\phi$-$N$ and a $p$-$n$ rescattering. Likewise, two distinct peaks can be seen for
$p_n=400~\mathrm{MeV}$: a $\phi$-$N$ rescattering peak at $\theta_{nl}=8^\circ$ and a
$p$-$n$ rescattering peak at $\theta_{nl}=65^\circ$, in agreement with the discussion of
Sec.~\ref{sec:geometry} (cf.~as well Fig.~\ref{fig:geometry}).

One new feature of the angular distributions in Fig.~\ref{fig:phicrx} is the presence of
a rescattering amplitude $\mathcal{M}_{1c}$ which corresponds to production of the $\phi$
meson from the neutron, which partially cancels the overall effect of the other FSIs since
it enters with a negative sign relative to $\mathcal{M}_{1a}$ and $\mathcal{M}_{1b}$. This
negative sign is due to antisymmetry of the deuteron wave-function under a
$p\leftrightarrow n$ transposition.

The presence of either screening or peaks at different recoil neutron momenta suggests
the ratio
\begin{equation}
	R = \frac{\sigma(p_n=400~\mathrm{MeV})}{\sigma(p_n=200~\mathrm{MeV})}
	\label{eqn:ratiophi}
\end{equation}
as especially conducive to our study, as the valleys in the angular distribution for
$p_n=200~\mathrm{MeV}$ entering into the denominator of $R$ will enhance the peaks
entering into the numerator.

\begin{figure}[hpt]
	\centering\includegraphics[scale=0.5]{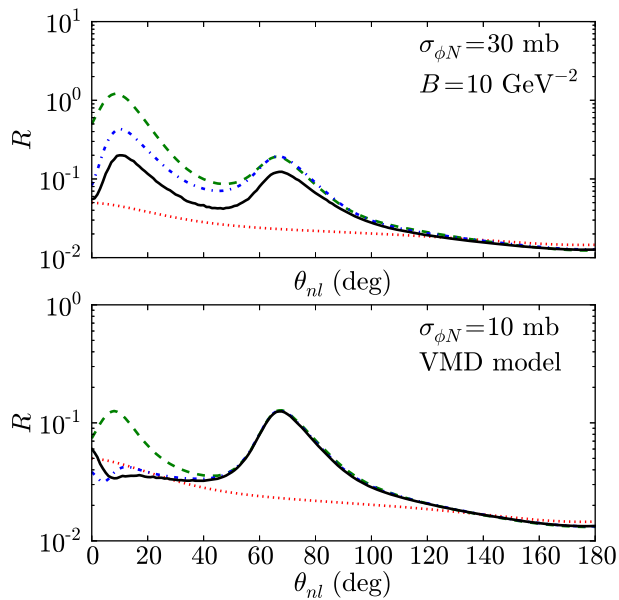}
	\caption{
		(Color online)
		Angular dependence of the ratio defined in Eq.~(\ref{eqn:ratiophi}),
		for two models of the $\phi$-$N$ interaction.
		Lines have the same meaning as in Fig.~\ref{fig:phicrx}.
		$q_0=5~\mathrm{GeV}$ and $-t=1.2~\mathrm{GeV}^2$.
	}
	\label{fig:phiratio}
\end{figure}

In Fig.~\ref{fig:phiratio} we present predictions for the ratio (\ref{eqn:ratiophi}) for
two models of $\phi$-$N$ scattering. In the upper panel, we use
$\sigma_{\phi N}=30~\mathrm{mb}$ and $B_{\phi N} = 10~\mathrm{GeV}^{-2}$ (the same
parameterization as in Fig.~\ref{fig:phicrx}), while in the lower panel we use
$\sigma_{\phi N}=10~\mathrm{mb}$ and assume vector meson dominance, so $\phi$-$N$
scattering has the same $t$ dependence as photoproduction. In both cases we use
$\alpha=-0.5$. As the figures show, the two models predict significantly different ratios
for the $\phi$-$N$ cross-section. Qualitatively, no $\phi$-$N$ rescattering peak is seen
for the $\sigma_{\phi N}=10~\mathrm{mb}$ model, as $\mathcal{M}_{1a}$ and
$\mathcal{M}_{1c}$ almost completely cancel. Quantitatively, the magnitudes of $R$ for the
two models differ by almost an order of magnitude at the $\phi$-$N$ rescattering peak.
Thus these estimates indicate that the break-up reaction (\ref{eqn:reaction}) is able to
effectively discriminate between different models of $\phi$-$N$ scattering, and can
complement other methods of studying the $\phi$-$N$ interaction.

\subsection{Photoproduction of $J/\Psi$ Mesons}
\label{sec:jpsi}

In studying the photoproduction of $J/\Psi$ in reaction (\ref{eqn:reaction}), we focus on
several aspects of the eikonal dynamics of the FSIs.  Namely, for the near-threshold
kinematics we will concentrate on identifying the $J/\Psi$-$N$ rescattering peak in the
angular distribution as we did in the case of $\phi$. For below-threshold production we
will elaborate the kinematic requirements of photoproduction, which result in a strong
suppression of FSI effects. Finally, for $J/\Psi$ production at collider energies, we
demonstrate almost complete cancelation of the $J/\Psi$-$N$ rescattering due to
destructive interference between single rescattering amplitudes.

\subsubsection{Near-threshold Photoproduction of $J/\Psi$ Mesons}

The upcoming $12~\mathrm{GeV}$ upgrade at Jefferson Lab\cite{Dudek:2012vr} will cross the
$J/\Psi$ photoproduction threshold for a proton target. This will provide unprecedented
opportunities for studies of $J/\Psi$ photoproduction at near-threshold kinematics, for
which there are currently very limited data (see
e.g.~Refs.~\cite{Gittelman:1975ix,Camerini:1975cy}).

This has renewed interest in theoretical studies of the $J/\Psi$-$N$ interaction at
near-threshold kinematics through photo- and electroproduction from the
deuteron\cite{Howell:2013fsa,Wu:2013xma}. The present analysis is focused on studying the
sensitivity of the reaction (\ref{eqn:reaction}) to the cross-section of the
$J/\Psi$-$N$ interaction through ratios similar to that of Eq.~(\ref{eqn:ratiophi}).

\begin{figure}[hpt]
	\centering\includegraphics[scale=0.5]{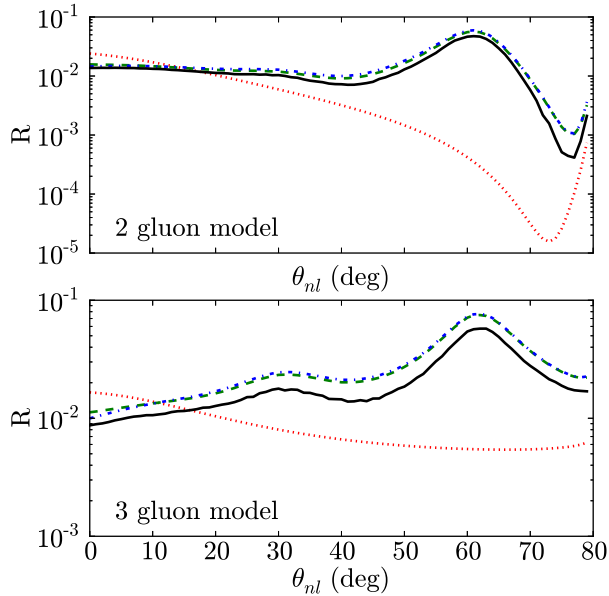}
	\caption{
		(Color online)
		Angular dependence of the ratio defined in Eq.~(\ref{eqn:ratiojpsi}), for the
		two- and three-gluon exchange models of $J/\Psi$ photoproduction.
		Assumes $\sigma_{\Psi N}=5~\mathrm{mb}$.
		Lines have the same meaning as in Fig.~\ref{fig:phicrx}.
		$q_0=10~\mathrm{GeV}$ and $t=t_{\mathrm{thr.}}=-2.23~\mathrm{GeV}^2$.
	}
	\label{fig:jpsi2g3g}
\end{figure}

The $J/\Psi$-nucleon interaction near threshold is not well understood. At high energies,
vector meson dominance suggests a total cross-section of around
$1~\mathrm{mb}$\cite{Knapp:1975nf}, while experimental data indicates a higher value of
$3.5~\mathrm{mb}$\cite{Anderson:1976hi}. On the theoretical side, the two-gluon exchange
models suggest a monotonically increasing total cross-section that approaches an
asymptotic value at large energies\cite{Kharzeev:1994pz}, and accordingly a smaller
cross-section near threshold. While these models are based on a leading twist
approximation, other models that attempt to account for nonperturbative effects
suggest a significantly larger cross section, as large as $17~\mathrm{mb}$ near
threshold\cite{Sibirtsev:2005ex}.

To ascertain the sensitivity of the considered break-up reaction to the $J/\Psi$-$N$
cross-section, we consider the ratio
\begin{equation}
	R = \frac{\sigma(p_n=600~\mathrm{MeV})}{\sigma(p_n=200~\mathrm{MeV})},
	\label{eqn:ratiojpsi}
\end{equation}
where for the numerator we choose the cross section at larger value of recoil nucleon
momenta to maximize the rescattering effects, and thereby the sensitivity of the $R$ to
the $J/\Psi$-$N$ scattering amplitude.

Since the $J/\Psi$ photoproduction amplitude is not factorized from the FSI amplitudes,
care should be given to the treatment of the $\gamma N\rightarrow J/\Psi N$ amplitude,
which can be strongly energy-dependent at near-threshold kinematics. For this reason we
consider two alternative models for $J/\Psi$ photoproduction near the threshold. In the
first model, the energy dependence of the photoproduction amplitude is estimated based on
the leading-twist two-gluon exchange model\cite{Brodsky:2000zc}, according to which the
function $A(s)$ entering into Eq.~(\ref{eqn:diffamp}) can be parameterized as
\begin{equation}
	A_{2g}(s) = \frac{\mathcal{N}_{2g}}{\sqrt{1+\alpha^2}}
		\left(\frac{s-s_{\mathrm{thr.},N}}{s-m_N^2}\right) 
	\label{eqn:jpsi2gA} ,
\end{equation}
with the constant $\mathcal{N}_{2g}=1.38~\mathrm{GeV}^{-2}$ found from a fit to
photoproduction data\cite{Camerini:1975cy}. The slope factor $B$ is estimated from the
two-gluon form factor of the nucleon as\cite{Frankfurt:2002ka}
\begin{equation}
	B_{\mathrm{eff.}} = \frac{4}{1~\mathrm{GeV}^2-t} .
\end{equation}
It is worth noting that at $t = -2.23~\mathrm{GeV}^2$,
$B_{\mathrm{eff.}}\approx1.24~\mathrm{GeV}^{-2}$, which is close to the low-energy $B$
value of $1.25~\mathrm{GeV}^{-2}$ measured at a Cornell experiment\cite{Gittelman:1975ix}.

The observation that the very limited phase space near threshold may favor a coherent 
interaction of all three quarks in the nucleon suggests the possible dominance of a
three-gluon exchange mechanism, which predicts much weaker energy dependence at the
threshold\cite{Brodsky:2000zc}:
\begin{equation}
	A_{3g}(s) = \frac{\mathcal{N}_{3g}}{\sqrt{1+\alpha^2}} ,
\end{equation}
with the constant $\mathcal{N}_{3g}=0.36~\mathrm{GeV}^{-2}$ fit to the results of
Ref.~\cite{Gittelman:1975ix}. For this model, we adopted a constant slope factor
$B = 1.25~GeV^{-2}$, consistent with experimental data\cite{Gittelman:1975ix}. In both
cases, we use $\alpha=-0.2$.

\begin{figure}[hpt]
	\centering\includegraphics[scale=0.5]{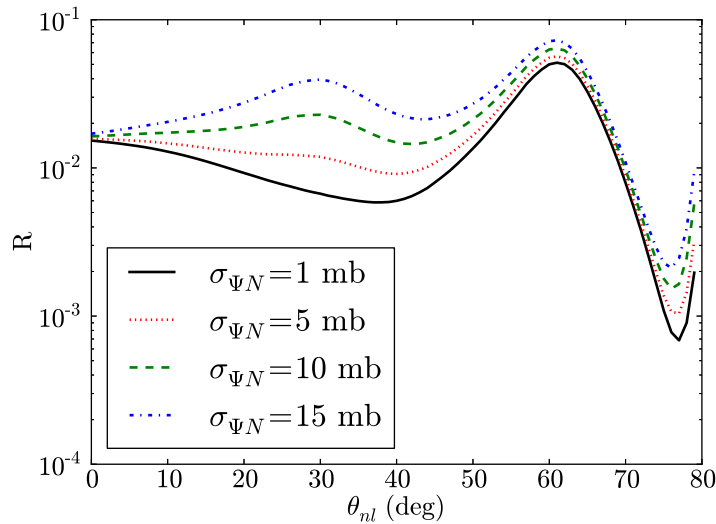}
	\caption{
		(Color online)
		Angular dependence of the ratio defined in Eq.~(\ref{eqn:ratiojpsi})
		with different $J/\Psi$-$N$ cross-sections,
		using the two-gluon parameterization of \cite{Brodsky:2000zc}.
		Estimated at $p_n=400~\mathrm{MeV}$,
		$q_0=10~\mathrm{GeV}$ and $-t=2.23~\mathrm{GeV}^2$.
		Includes all diagrams.
	}
	\label{fig:jpsimulti}
\end{figure}

In Fig.~\ref{fig:jpsi2g3g} we present the prediction of above models, where the
diffractive part of the $J\Psi$-$N$ amplitude is parameterized with
$\sigma_{\Psi N}=5~\mathrm{mb}$ and $B = 1.25~\mathrm{GeV}^{-2}$. This figure indicates
the presence of two rescattering peaks, with the peak at $\theta_{nl}=30^\circ$
corresponding to $J/\Psi$-$N$ rescattering. This pattern of two rescattering peaks, each
corresponding to one of the rescattering diagrams, has also been observed in
Ref.~\cite{Howell:2013fsa}.

These estimates also demonstrate the sensitivity of the reaction (\ref{eqn:reaction})
to the energy dependence of the photoproduction amplitude. For the two-gluon mechanism,
a dip should be present near the kinematical limit of $\theta_{nl}$ due to the threshold
factor of $s-s^0_{\mathrm{thr.},N}$. At this $\theta_{nl}$, the angle between the outgoing
proton and the $J/\Psi$ meson becomes smaller, thus making
$s_{\gamma p} = m_N^2+m_\Psi^2+2E_pE_\Psi-2\mathbf{p}_p\cdot\mathbf{p}_\Psi$
smaller as well. No such suppression factor exists for the three-gluon model. Besides
this, Fig.~\ref{fig:jpsi2g3g} demonstrates that the effect of $\mathcal{M}_{1c}$ is
negligible, owing to the large value of $-t_{\mathrm{thr.}}$.

We conclude our discussion of near-threshold $J/\Psi$ production from the deuteron by
presenting, in Fig.~\ref{fig:jpsimulti}, estimates of $R$ (see Eq.~\ref{eqn:ratiojpsi})
for the two-gluon exchange model of Eq.~(\ref{eqn:jpsi2gA}) for several values of
$\sigma_{\Psi N}$, thus demonstrating the sensitivity of reaction (\ref{eqn:reaction}) to
the total $J/\Psi$-$N$ cross-section.

\subsubsection{Sub-threshold Photoproduction of $J/\Psi$ Mesons}

\begin{figure}[hpt]
	\centering\includegraphics[scale=0.5]{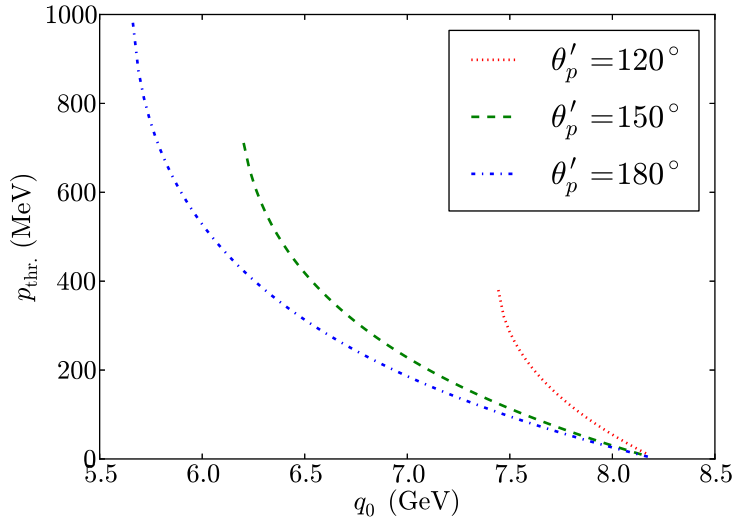}
	\caption{
		(Color online)
		The threshold value for $J/\Psi$ photoproduction of the bound proton momentum as
		a function of photon energy for various proton-photon angles.
	}
	\label{fig:subkine}
\end{figure}

While the threshold for $J/\Psi$ photoproduction from a stationary proton is
$E_{\mathrm{thr.},N}=8.2~\mathrm{GeV}$, the threshold for production from the deuteron is
only $5.6~\mathrm{GeV}$, making it kinematically possible to produce $J/\Psi$ with
$q_0<E_{\mathrm{thr.},N}$. This requires that the bound nucleon struck by the photon be
sufficiently fast so that $s_{\gamma N} \geq s_{\mathrm{thr.},N} = (m_n+m_\Psi)^2$. The
kinematics of reaction (\ref{eqn:reaction}) allow for this when the recoil neutron flies
off in the forward direction. Within PWIA this will correspond to the $J/\Psi$ meson off
a backward moving proton that satisfy the condition
\begin{equation}
	(q + p_p^\prime)^2 > s_{\mathrm{thr.},N}
	\label{eqn:sthr} .
\end{equation}
This condition can only be satisfied for fast, backwards-going protons that satisfy the
condition
\begin{equation}
	|\mathbf{p}_p^\prime| \geq p_{\mathrm{thr.}}(q_0,\theta_p^\prime)
	\label{eqn:pthr} ,
\end{equation}
where $\theta_p^\prime$ is the angle between the bound proton and the photon. This
threshold momentum is shown as a function of $q_0$ in Fig.~\ref{fig:subkine} for several
values of $\theta_p^\prime$. The value of $p_{\mathrm{thr.}}$ can be very large for
small photon energies, even above $700~\mathrm{MeV}$. At these photon energies, we expect
the applicability of VNA to break down, so its predictions will be rather qualitative.
However, key features such as the onset of the eikonal regime for rescattering
will be relevant to other (e.g. light cone) approximations which are better suited to
describing sub-threshold production.

\begin{figure}[hpt]
	\centering\includegraphics[scale=0.5]{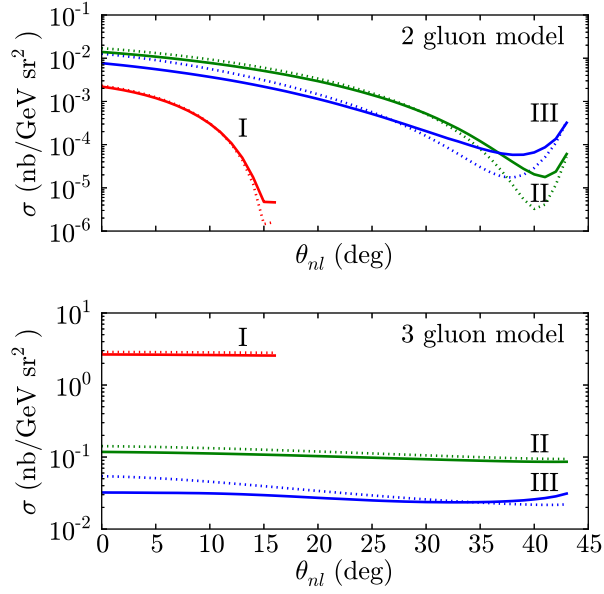}
	\caption{
		(Color online)
		Angular dependence of the $J/\Psi$ photoproduction cross-section at different
		neutron momenta below threshold.
		Assumes $\sigma_{\Psi N}=5~\mathrm{mb}$.
		The curves I, II, and III represent $p_n$ equal to $200$, $400$, and $600$ MeV,
		respectively.
		The dotted and solid lines are PWIA and full calculations, respectively.
		$q_0=7~\mathrm{GeV}$ and $t=t_{\mathrm{thr.},N}=-2.23~\mathrm{GeV}^2$.
	}
	\label{fig:jpsisub}
\end{figure}

As for FSIs, we may expect immediately that $\mathcal{M}_{1c}$ does not contribute, since
$-t_{\mathrm{thr.}}$ is large and $q_0$ is small. As for $\mathcal{M}_{1a}$ and
$\mathcal{M}_{1b}$, one of the analytic properties of their pole parts
(Eqs.~(\ref{eqn:M1apole},\ref{eqn:M1bpole})) is that
\begin{equation}
	p_{n,z} = p_{n,z}^\prime + \Delta_{1a/1b}
\end{equation}
which, because of momentum conservation at the $d\rightarrow pn$ vertex, implies
\begin{equation}
	p_{n,z} = -p_{p,z}^\prime + \Delta_{1a/ab/1c}
		= -|\mathbf{p}_p^\prime|\cos(\theta_p^\prime) + \Delta_{1a/1b} .
	\label{eqn:subdelta}
\end{equation}
As can be seen in Eqs.~(\ref{eqn:delta1a},\ref{eqn:delta1b}), the factors $\Delta_{1a}$
and $\Delta_{1b}$ are both positive. This means that
$p_{n,z} > -|\mathbf{p}_p^\prime|\cos(\theta_p^\prime)$, with the lesser quantity a
positive number since the struck proton will be backwards-going. To be compatible with the
inequality (\ref{eqn:pthr}), this requires an especially large $p_{n,z}$. However, the
$\Delta$ factors monotonically increase with $p_{n,z}$, making the condition
(\ref{eqn:subdelta}) more difficult to fulfill. Thus, we expect a suppression of FSIs for
sub-threshold kinematics. This is illustrated in Fig.~\ref{fig:jpsisub}, where we present
numerical estimates for the $J/\Psi$ production cross-section at $q_0=7~\mathrm{GeV}$
using the two- and three-gluon exchange models of Ref.~\cite{Brodsky:2000zc}. As the
figure shows, the FSI effects are small and, contrary to above threshold kinematics, do
not increase with increased $p_n$. Looking back at Fig.~\ref{fig:subkine}, we conclude
that at sub-threshold kinematics the reaction (\ref{eqn:reaction}) in the limit
$q_0\rightarrow E_{\mathrm{thr.}}^{\mathrm{deut.}}$ will allow probing of the internal
structure of the deuteron at large Fermi momenta with little distortion from FSIs.

\subsection{Photoproduction of $J/\Psi$ at Collider Energies}

As was mentioned in Sec.~\ref{sec:singles}, due to antisymmetry of deuteron wave-function, 
the $J/\Psi$-$N$ rescatteirng amplitude of Fig.~\ref{fig:resc1}({c}) enters with the
opposite sign relative to the other $J/\Psi$-$N$ rescattering amplitude of
Fig.~\ref{fig:resc1}({a}). It was observed in Sec.~\ref{sec:jpsi} that the contribution of
$\mathcal{M}_{1c}$ will be small for near-threshold kinematics, so there is little
cancelation with the contribution of $\mathcal{M}_{1a}$. This was due to the large value
of $t_{\mathrm{min.}}\sim t_{\mathrm{thr.}}$, which required both vertices in
Fig.~\ref{fig:resc1}({c}) to be hard.

However, for large photon energies, $t_{\mathrm{min.}}\rightarrow 0$ and thus we expect
almost complete cancelation between $\mathcal{M}_{1a}$ and $\mathcal{M}_{1c}$. This
represents a situation where the $J/\Psi$ meson will not participate in FSIs.

\begin{figure}[hpt]
	\centering\includegraphics[scale=0.5]{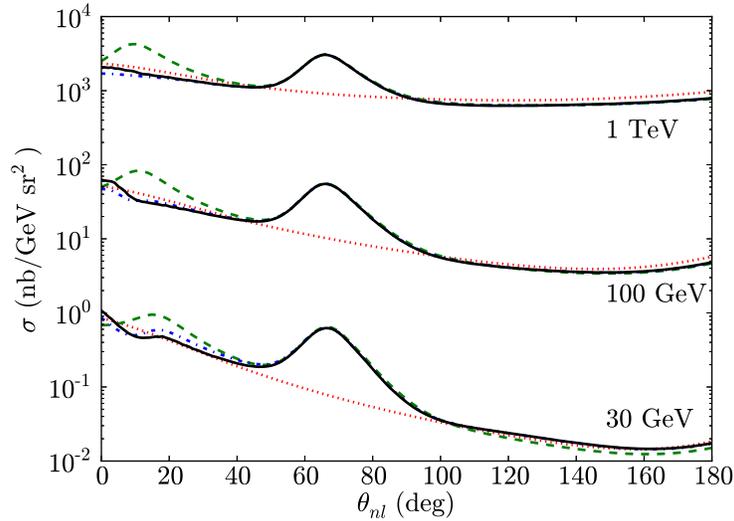}
	\caption{
		(Color online)
		Angular dependence of the $J/\Psi$ photoproduction cross-section at collider
		energies, given $t=-1.5~\mathrm{GeV}^2$ and $p_n=400~\mathrm{MeV}$.
		Assumes $\sigma_{\Psi N}=5~\mathrm{mb}$.
		Lines have the same meaning as in Fig.~\ref{fig:phicrx}.
	}
	\label{fig:jpsicollider}
\end{figure}

To estimate the cancelation effect numerically, we calculated the cross-section of the
reaction (\ref{eqn:reaction}) for collider energies. We parameterize the $s$ dependence of
the $\gamma N\rightarrow J/\Psi N$ amplitude using a leading-order pQCD
approximation\cite{Rebyakova:2011vf}. In particular, for the function $A(s)$ entering into
Eq.~(\ref{eqn:diffamp}), we use
\begin{equation}
	A(s) = \mathcal{N} \alpha_s(\mu^2) x G_T(x,\mu^2) ,
\end{equation}
where $G_T(x,\mu^2)$ is the gluonic distribution of the target nucleon at factorization
scale $\mu^2$. Here $x = \frac{m_\Psi^2}{s_{\gamma N^*}}$, and we choose
$\mu^2 = \frac{m_\Psi^2}{4}$ similar to \cite{Rebyakova:2011vf}. For these estimates
we used the CJ12 next-to-leading order partonic distribution functions\cite{Owens:2012bv},
and we modeled the $t$ dependence using a slope factor of
$B=4.73~\mathrm{GeV}^{-2}$\cite{Adloff:2000vm}. 

The results for three photon beam energies (with a recoil neutron momentum of
$400~\mathrm{MeV}$) are given in Fig.~\ref{fig:jpsicollider}. They demonstrate significant
cancelation of the two $J/\Psi$-$N$ rescattering amplitudes, with the cancelation becoming
predominant with increasing photon energy. Even for $q_0=30~\mathrm{GeV}$ the FSI
effects at $\theta_{nl}\approx 30^\circ$ (where one expects maximal $J/\Psi$-$N$
rescattering) contribute less than a few percent of the total cross-section. The overall
FSI effects are due only to $p$-$n$ reinteractions.

It is worth noting that the observed cancelation is also valid for other nuclei, since it
is due to the antisymmetry of the nuclear wave-function. Thus the present result suggests
that $J/\Psi$ mesons produced from nuclei in coincidence with a knock-out nucleon will not
undergo first order $J/\Psi$-$N$ rescattering, which may significantly reduce their
absorption.

\section{Conclusions}
\label{sec:conclusion}

With the virtual nucleon approximation, we calculated the large $-t\ge 1~\mathrm{GeV}^2$
cross-section for vector meson photoproduction from the deuteron accompanied by deuteron
break-up. We focused on kinematics where one of the nucleons could be identified as struck
and the other as a spectator to the $\gamma N\rightarrow VN$ subprocess. The large $-t$
involved validates the generalized eikonal approximation in calculating the final state
interactions.

Our results indicate that, similarly to the $\gamma^* d\rightarrow pn$ reaction, one can
observe a minimum or peak in the recoil nucleon angular distribution associated with a
$p$-$n$ reinteraction. However, a novelty of the photoproduction reaction is that a second
mininum or peak appears due to $V$-$N$ rescattering. Using $\phi(1020)$ and $J/\Psi$ as
examples, we showed that this second peak (or minimum) can be used to probe
characteristics of the largely unknown $V$-$N$ interaction. In the case of $J/\Psi$ in
particular, we extended our calculations into the sub-threshold and high-energy domains.

The sub-threshold calculations demonstrated and overall suppression of final state
interactions due to restrictive kinematical limitations. This result indicates that
sub-threshold production can be used to probe the deuteron at extremely large internal
momenta with little distortion due to FSIs.

In the high-energy limit, we observe a significant cancelation between $J/\Psi$-$N$
rescattering amplitudes associated with production from different nucleons in the
deuteron. In this limit, FSI effects are due entirely to $p$-$n$ rescattering. This result
suggests experiments where the $J/\Psi$ meson is detected in coincidence with a knocked
out nucleon as a means of probing the $J/\Psi$ photoproduction with little contribution
from $J/\Psi$-$N$ rescattering, as the cancelation strongly suppresses this contribution.
 
\section{Acknowledgements}

We are thankful to Drs. Gerald Miller, Mark Strikman, and Xin Qian for helpful comments
and  discussions. This work is supported by the U.S. Department of Energy grant under
contract DE-FG02-01ER41172.

\appendix*
\section{Double Rescattering Term}

Here we give the details of the derivation of the double scattering amplitudes presented
in Sec.~\ref{sec:doubles}. The derivation follows these steps of GEA:
\begin{enumerate}
	\item The full amplitude is written out using the effective Feynman rules.
	\item Energy integration places the spectator to photoproduction and a
		low-momentum internal nucleon on its mass shell.
	\item Terms are gathered into the VNA deuteron wave-function.
	\item A decomposition into pole and principal value parts is performed for each
		remaining propagator.
	\item Completeness relations are used and terms gathered into subprocess
		(photoproduction and reinteraction) amplitudes.
\end{enumerate}
All of the steps shall be here carried out in full detail for $\mathcal{M}_{2a}$, whereas
a more cursory description shall be given for $\mathcal{M}_{2b}$ and $\mathcal{M}_{2c}$.

To begin, the full amplitude for the process in Fig. $\ref{fig:resc2}(a)$ is
\begin{align}
	\mathcal{M}_{2a}^{(\lambda_V,\lambda_p,\lambda_n;\lambda_\gamma,\lambda_d)}
		&=
		-
		\bar{u}^{(\lambda_p)}(\mathbf{p}_p)
		\bar{u}^{(\lambda_n)}(\mathbf{p}_n)
		{\phi^*_\pi}^{(\lambda_V)}(\mathbf{p}_V)
		\int\frac{d^4p_n^\prime}{(2\pi)^4i}
		\int\frac{d^4p_n^{\prime\prime}}{(2\pi)^4i}
		\bigg[
		\Gamma_{pn\rightarrow pn}
		\frac{\slashed{p}_p^{\prime\prime}+m_N}{(p_p^{\prime\prime})^2-m_N^2+i\epsilon}
	\notag \\ &\times
		\frac{\slashed{p}_n^{\prime\prime}+m_N}{(p_n^{\prime\prime})^2-m_N^2+i\epsilon}
		\Gamma^{\rho\pi}_{VN\rightarrow VN}
		\frac{G_{\rho\nu}(p_V^\prime)}{(p_V^\prime)^2-m_V^2+i\epsilon}
		\Gamma^{\mu\nu}_{\gamma N\rightarrow VN}
		\epsilon^{(\lambda_\gamma)}_\mu(q)
		\frac{\slashed{p}_p^\prime+m_N}{(p_p^\prime)^2-m_N^2+i\epsilon}
	\notag \\ &\times
		\frac{\slashed{p}_n^\prime+m_N}{(p_n^\prime)^2-m_N^2+i\epsilon}
		\Gamma_{dpn}
		\chi^{(\lambda_d)}
		\bigg]
	\label{eqn:M2afull} ,
\end{align}
where we have chosen the integration variable in anticipation of the positive-energy,
on-shell projections that shall be executed. Firstly, as prescribed by VNA, a slow nucleon
with momentum $p_n^\prime$ shall be placed on its mass shell. Of the remaining internal
lines, $p_n^{\prime\prime}$ has the smallest momentum, since the photoproduction vertex is
hard, giving the proton a large momentum, whereas the $V$-$N$ rescattering is soft. Doing
these energy integrations and keeping the positive-energy parts, as well as using the
exact completeness relations entailed by placing the neutron on-shell, gives
\begin{align}
	\mathcal{M}_{2a}^{(\lambda_V,\lambda_p,\lambda_n;\lambda_\gamma,\lambda_d)}
		&=
		-
		\bar{u}^{(\lambda_p)}(\mathbf{p}_p)
		\bar{u}^{(\lambda_n)}(\mathbf{p}_n)
		{\phi^*_\pi}^{(\lambda_V)}(\mathbf{p}_V)
		\sum_{\lambda_n^\prime\lambda_n^{\prime\prime}}
		\int\frac{d^3\mathbf{p}_n^\prime}{(2\pi)^3}
		\int\frac{d^3\mathbf{p}_n^{\prime\prime}}{(2\pi)^3}
		\bigg[
		\frac{1}{2E_n^\prime}
		\frac{1}{2E_n^{\prime\prime}}
	\notag \\ &\times
		\Gamma_{pn\rightarrow pn}
		u^{(\lambda_n^{\prime\prime})}(\mathbf{p}_n^{\prime\prime})
		\bar{u}^{(\lambda_n^{\prime\prime})}(\mathbf{p}_n^{\prime\prime})
		\frac{\slashed{p}_p^{\prime\prime}+m_N}{(p_p^{\prime\prime})^2-m_N^2+i\epsilon}
		\Gamma^{\rho\pi}_{VN\rightarrow VN}
		\frac{G_{\rho\nu}(p_V^\prime)}{(p_V^\prime)^2-m_V^2+i\epsilon}
	\notag \\ &\times
		u^{(\lambda_n^\prime)}(\mathbf{p}_n^\prime)
		\Gamma^{\mu\nu}_{\gamma N\rightarrow VN}
		\epsilon^{(\lambda_\gamma)}_\mu(q)
		\frac{\slashed{p}_p^\prime+m_N}{(p_p^\prime)^2-m_N^2+i\epsilon}
		\bar{u}^{(\lambda_n^\prime)}(\mathbf{p}_n^\prime)
		\Gamma_{dpn}
		\chi^{(\lambda_d)}
		\bigg]
	\label{eqn:M2aEnergy}
	.
\end{align}
Next, we use the approximate completeness relation
$\slashed{p}_p^\prime+m_N\approx\sum_{\lambda_p^\prime}
	u^{(\lambda_p^\prime)}(\mathbf{p}_p^\prime)
	\bar{u}^{(\lambda_p^\prime)}(\mathbf{p}_p^\prime)$
for the virtual, struck proton so that we may gather terms into the deuteron
wave-function defined in Eq.~(\ref{eqn:wfd}), as follows:
\begin{align}
	\mathcal{M}_{2a}^{(\lambda_V,\lambda_p,\lambda_n;\lambda_\gamma,\lambda_d)}
		&=
		-
		\bar{u}^{(\lambda_p)}(\mathbf{p}_p)
		\bar{u}^{(\lambda_n)}(\mathbf{p}_n)
		{\phi^*_\pi}^{(\lambda_V)}(\mathbf{p}_V)
		\sum_{\lambda_p^\prime\lambda_n^\prime\lambda_n^{\prime\prime}}
		\int\frac{d^3\mathbf{p}_n^\prime}{(2\pi)^3}
		\int\frac{d^3\mathbf{p}_n^{\prime\prime}}{(2\pi)^3}
		\bigg[
		\sqrt{\frac{2(2\pi)^3}{2E_n^\prime}}
		\frac{1}{2E_n^{\prime\prime}}
	\notag \\ &\times
		\Gamma_{pn\rightarrow pn}
		u^{(\lambda_n^{\prime\prime})}(\mathbf{p}_n^{\prime\prime})
		\bar{u}^{(\lambda_n^{\prime\prime})}(\mathbf{p}_n^{\prime\prime})
		\frac{\slashed{p}_p^{\prime\prime}+m_N}{(p_p^{\prime\prime})^2-m_N^2+i\epsilon}
		\Gamma^{\rho\pi}_{VN\rightarrow VN}
		\frac{G_{\rho\nu}(p_V^\prime)}{(p_V^\prime)^2-m_V^2+i\epsilon}
	\notag \\ &\times
		u^{(\lambda_n^\prime)}(\mathbf{p}_n^\prime)
		\Gamma^{\mu\nu}_{\gamma N\rightarrow VN}
		\epsilon^{(\lambda_\gamma)}_\mu(q)
		u^{(\lambda_p^\prime)}(\mathbf{p}_p^\prime)
		\Psi_d^{(\lambda_p^\prime,\lambda_n^\prime;\lambda_d)}(\mathbf{p}_n^\prime)
		\bigg]
	\label{eqn:M2adwf}
	.
\end{align}
From here, we change the integration variables to the transferred momentum at each vertex,
namely
\begin{align}
	K &= p_n-p_n^{\prime\prime} \\
	K^\prime &= p_n^{\prime\prime}-p_n^\prime
\end{align}
and write the remaining propagators in terms of these momenta. For the proton propagator,
\begin{align}
	(p_p^{\prime\prime})^2-m_N^2+i\epsilon
		&= (p_p+p_n-p_n^{\prime\prime})^2-m_N^2+i\epsilon
		= (p_p+K)^2-m_N^2+i\epsilon
		\notag \\ &= K^2+2E_pK_0-\mathbf{p}_p\cdot\mathbf{K}_\perp-p_{p,z}K_z+i\epsilon
		\notag \\ &= 2p_{p,z}(\Delta_{2a}-K_z+i\epsilon) ,
\end{align}
where
\begin{equation}
	\Delta_{2a} = \frac{K^2+2E_pK_0-2\mathbf{p}_p\cdot\mathbf{K}_\perp}{2p_{p,z}}
\end{equation}
and for the vector meson propagator
\begin{align}
	(p_V^\prime)^2-m_V^2+i\epsilon
		&= (p_V+p_n^{\prime\prime}-p_n^\prime)^2-m_V^2+i\epsilon
		= (p_V+K^\prime)^2-m_V^2+i\epsilon
		\notag \\ &= (K^\prime)^2+2E_VK^\prime_0-\mathbf{p}_V\cdot\mathbf{K}^\prime_\perp-p_{V,z}K^\prime_z+i\epsilon
		\notag \\ &= 2p_{V,z}(\Delta^\prime_{2a}-K^\prime_z+i\epsilon) ,
\end{align}
where
\begin{equation}
	\Delta^\prime_{2a} = \frac{(K^\prime)^2+2E_VK^\prime_0-2\mathbf{p}_V\cdot\mathbf{K}^\prime_\perp}{2p_{V,z}} .
\end{equation}
With the propagator denominators in this form, we can decompose the integrals over each of
$K_z$ and $K^\prime_z$ into pole and principal value parts. Because the principal value
part will be small compared to the pole part, and because double rescattering will itself
be small compared to single rescattering, we consider only the pole parts of each
integral, placing both the proton and vector meson on their mass shells. This allows us to
use completeness relations for both, giving
%(after some terms have been shuffled around)
\begin{align}
	\mathcal{M}_{2a}^{(\lambda_V,\lambda_p,\lambda_n;\lambda_\gamma,\lambda_d)}
		&=
		-\frac{1}{4p_{p,z}p_{V,z}}
		\sum_{\lambda_p^\prime\lambda_n^\prime\lambda_p^{\prime\prime}\lambda_n^{\prime\prime}\lambda_V^\prime}
		\int\frac{d^2\mathbf{K}_\perp}{(2\pi)^2}
		\int\frac{d^2\mathbf{K}^\prime_\perp}{(2\pi)^2}
		\bigg[
		\sqrt{\frac{2(2\pi)^3}{2E_n^\prime}}
		\frac{1}{2E_n^{\prime\prime}}
	\notag \\ &\times
		\bar{u}^{(\lambda_p)}(\mathbf{p}_p)
		\bar{u}^{(\lambda_n)}(\mathbf{p}_n)
		\Gamma_{pn\rightarrow pn}
		u^{(\lambda_n^{\prime\prime})}(\mathbf{p}_n^{\prime\prime})
		u^{(\lambda_p^{\prime\prime})}(\mathbf{p}_p^{\prime\prime})
	\notag \\ &\times
		\bar{u}^{(\lambda_n^{\prime\prime})}(\mathbf{p}_n^{\prime\prime})
		{\phi^*_\pi}^{(\lambda_V)}(\mathbf{p}_V)
		\Gamma^{\rho\pi}_{VN\rightarrow VN}
		{\phi^*_\nu}^{(\lambda_V^\prime)}(\mathbf{p}_V^\prime)
		u^{(\lambda_n^\prime)}(\mathbf{p}_n^\prime)
	\notag \\ &\times
		\bar{u}^{(\lambda_p^{\prime\prime})}(\mathbf{p}_p^{\prime\prime})
		\phi^{(\lambda_V^\prime)}_\rho(\mathbf{p}_V^\prime)
		\Gamma^{\mu\nu}_{\gamma N\rightarrow VN}
		\epsilon^{(\lambda_\gamma)}_\mu(q)
		u^{(\lambda_p^\prime)}(\mathbf{p}_p^\prime)
	\notag \\ &\times
		\Psi_d^{(\lambda_p^\prime,\lambda_n^\prime;\lambda_d)}
			(p_{n,z}-\Delta_{2a}-\Delta_{2a}^\prime;\mathbf{p}_{n,\perp}-\mathbf{K}_\perp-\mathbf{K}^\prime_\perp)
		\bigg]
	\label{eqn:M2acompl}
	.
\end{align}
Terms are then gathered into subprocess (photoproduction and rescattering) amplitudes,
giving
\begin{align}
	\mathcal{M}_{2a}^{(\lambda_V,\lambda_p,\lambda_n;\lambda_\gamma,\lambda_d)}
		&=
		-\frac{1}{4p_{p,z}p_{V,z}}
		\sum_{\mathrm{spins}}
		\int\frac{d^2\mathbf{K}}{(2\pi)^2}
		\int\frac{d^2\mathbf{K}^\prime}{(2\pi)^2}
		\bigg[
		\sqrt{\frac{2(2\pi)^3}{2E_n^\prime}}
		\frac{1}{2E_n^{\prime\prime}}
		\mathcal{M}_{pn\rightarrow pn}^{(\lambda_p,\lambda_n;\lambda_p^{\prime\prime},\lambda_n^{\prime\prime})}
			(p_p,p_n;p_p^{\prime\prime},p_n^{\prime\prime})
	\notag \\ &\times
		\mathcal{M}_{VN\rightarrow VN}^{(\lambda_V,\lambda_n^{\prime\prime};\lambda_V^\prime,\lambda_n^\prime)}
			(p_V,p_n^{\prime\prime};p_V^\prime,p_n^\prime)
		\mathcal{M}_{\gamma N\rightarrow VN}^{(\lambda_V^\prime,\lambda_p^{\prime\prime};\lambda_\gamma,\lambda_p^\prime)}
			(p_V^\prime,p_p^{\prime\prime};p_\gamma,p_p^\prime)
	\notag \\ &\times
		\Psi_d^{(\lambda_p^\prime,\lambda_n^\prime;\lambda_d)}
			(p_{n,z}-\Delta_{2a}-\Delta_{2a}^\prime;\mathbf{p}_{n,\perp}-\mathbf{K}_\perp-\mathbf{K}^\prime_\perp)
		\bigg]
	\label{eqn:M2afinal} ,
\end{align}
where the sum is over all internal spins. This is exactly Eq.~(\ref{eqn:M2a}).

The same process is followed for calculating $\mathcal{M}_{2b}$ and $\mathcal{M}_{2c}$,
but with a few minor differences. First, regarding $\mathcal{M}_{2b}$. The momentum line
$p_n^\prime$ corresponds to the spectator, so it is to be taken on-shell via energy
integration. Of the remaining momentum lines, $p_n^{\prime\prime}$ is slow, so it too is
placed on-shell by energy integration. The propagator for $p_p^\prime$ is absorbed into
the deuteron wave-function, so the remaining propagators are for $p_p^{\prime\prime}$ and
$p_V^\prime$. Momentum conservation, as before, can be used to rewrite the propagators in
terms of the transferred momenta $K=p_n-p_n^{\prime\prime}$ and
$K^\prime=p_n^{\prime\prime}-p_n^\prime$ and to find what the delta factors are.
In particular,
\begin{align}
	(p_V^\prime)^2-m_V^2+i\epsilon
		&= (p_V+p_n-p_n^{\prime\prime})^2-m_V^2+i\epsilon
		= (p_V+K)^2-m_V^2+i\epsilon
		\notag \\ &= K^2+2E_VK_0-\mathbf{p}_V\cdot\mathbf{K}_\perp-p_{V,z}K_z+i\epsilon
		\notag \\ &= 2p_{V,z}(\Delta_{2a}-K_z+i\epsilon)
	\\
	(p_p^{\prime\prime})^2-m_N^2+i\epsilon
		&= (p_p+p_n^{\prime\prime}-p_n^\prime)^2-m_N^2+i\epsilon
		= (p_p+K^\prime)^2-m_N^2+i\epsilon
		\notag \\ &= (K^\prime)^2+2E_pK^\prime_0-\mathbf{p}_p\cdot\mathbf{K}^\prime_\perp-p_{p,z}K^\prime_z+i\epsilon
		\notag \\ &= 2p_{p,z}(\Delta^\prime_{2b}-K^\prime_z+i\epsilon) ,
\end{align}
where
\begin{align}
	\Delta_{2b} &= \frac{K^2+2E_VK_0-2\mathbf{p}_V\cdot\mathbf{K}_\perp}{2p_{V,z}} \\
	\Delta^\prime_{2b} &= \frac{(K^\prime)^2+2E_pK^\prime_0-2\mathbf{p}_p\cdot\mathbf{K}^\prime_\perp}{2p_{p,z}} .
\end{align}
Next, regarding $\mathcal{M}_{2c}$. The momentum line $p_p^\prime$ is the spectator to
photoproduction, so it is placed on-shell by energy integration. The $V$-$N$ rescattering
vertex is hard, so $p_p^{\prime\prime}$ is fast, whereas the photoproduction vertex is
soft, making $p_n^{\prime\prime}$ slow. Thus, it is $p_n^{\prime\prime}$ that will also be
put on its mass shell. The propagator for $p_n^\prime$ is absorbed into the deuteron
wave-function, so the remaining propgators are those for $p_V^\prime$ and
$p_p^{\prime\prime}$. By momentum conservation, we rewrite them in the following forms and
find the delta factors:
\begin{align}
	(p_p^{\prime\prime})^2-m_N^2+i\epsilon
		&= (p_p+p_n-p_n^{\prime\prime})^2-m_N^2+i\epsilon
		= (p_p+K)^2-m_N^2+i\epsilon
		\notag \\ &= K^2+2E_pK_0-\mathbf{p}_p\cdot\mathbf{K}_\perp-p_{p,z}K_z+i\epsilon
		\notag \\ &= 2p_{p,z}(\Delta_{2c}-K_z+i\epsilon)
	\\
	(p_V^\prime)^2-m_V^2+i\epsilon
		&= (q+p_n^{\prime\prime}-p_n^\prime)^2-m_V^2+i\epsilon
		= (q-K^\prime)^2-m_V^2+i\epsilon
		\notag \\ &= (K^\prime)^2-2q_0K^\prime_0-q_0K^\prime_z+i\epsilon
		\notag \\ &= 2q_0(K^\prime_z-\Delta^\prime_{2c}+i\epsilon) ,
\end{align}
where
\begin{align}
	\Delta_{2c} &= \frac{K^2+2E_pK_0-2\mathbf{p}_p\cdot\mathbf{K}_\perp}{2p_{p,z}} \\
	\Delta^\prime_{2c} &= \frac{(K^\prime)^2-m_V^2}{2q_0}+K^\prime_0 .
\end{align}
After this procedure is done for $\mathcal{M}_{2b}$ and $\mathcal{M}_{2c}$, the pole term
in each integral is kept, completeness relations are used, and terms are gathered into
subprocess amplitudes, resulting in Eqs. (\ref{eqn:M2b}) and (\ref{eqn:M2c}).

\bibliographystyle{apsrev4-1}
\bibliography{references}

\end{document}